\documentclass[pre,aps,amsmath,twocolumn,showpacs,superscriptaddress,
longbibliography]{revtex4-1}

\usepackage{amssymb}
\usepackage{graphicx}
\usepackage[pdftex,bookmarks,colorlinks,breaklinks]{hyperref}
\hypersetup{linkcolor=red,citecolor=blue,filecolor=dullmagenta,urlcolor=blue}
\usepackage{amsmath}
\usepackage{mathrsfs}
\usepackage{amsfonts}

\begin{document}

\title{Universal insulating--to--metallic crossover in tight-binding random geometric graphs}

\author{A.~M. Mart\'inez-Arg\"uello}
\email{blitzkriegheinkel@gmail.com}
\affiliation{Instituto de F\'isica, Benem\'erita Universidad Aut\'onoma de Puebla, Puebla 72570, Mexico}

\author{K.~B. Hidalgo-Castro}
\affiliation{Instituto de F\'isica, Benem\'erita Universidad Aut\'onoma de Puebla, Puebla 72570, Mexico}

\author{J.~A. M\'endez-Berm\'udez}
\affiliation{Instituto de F\'isica, Benem\'erita Universidad Aut\'onoma de Puebla, Puebla 72570, Mexico}


\begin{abstract}

Within the scattering matrix approach to electronic transport, the scattering and transport properties of tight-binding random graphs are analyzed. In particular, we compute the scattering matrix elements, the transmission, the channel-to-channel transmission distributions (including the total transmission distribution), the shot noise power, and the elastic enhancement factor. Two graph models are considered: random geometric graphs and bipartite random geometric graphs. The results show an insulating to a metallic crossover in the scattering and transport properties by increasing the average degree of the graphs from small to large values. Also, the scattering and transport properties are shown to be invariant under a scaling parameter depending on the average degree and the graph size. Furthermore, for large connectivity and in the perfect coupling regime, the scattering and transport properties of both graph models are well described by the random matrix theory predictions of electronic transport, except for bipartite graphs in particular scattering setups.

\end{abstract}

\pacs{46.65.+g, 89.75.Hc, 05.60.Gg}

\maketitle


\section{Introduction}

Random networks or random graphs have been of great interest in the last decades (see for instance Refs.~\cite{Strogatz2001,Dorogovtsev2001,Albert2002,Newman2003,Boccalettia2006,DurretBook,LatoraBook,EstradaBook,PenroseBook} and the references therein). This is due to the fact that they have proved to be a very valuable tool for the description of complex systems appearing in research areas ranging from mathematics~\cite{Erdos1959,Chung1997}, physics~\cite{Bianconi2001,Dorogovtsev2008,Estrada2012,Holovatch2017,Biamonte2019,Cimini2019,Jusup2022}, chemistry~\cite{Lehn2013,Rappoport2014,Ozkanlar2014} and biology~\cite{Dorogovtsev2008,Estrada2012,Kohn1999,Hartwell1999,Bhalla1999} to economics~\cite{Jusup2022} and social sciences~\cite{Wong2006,Jusup2022}, including many others~\cite{Jusup2022,Barabasi1999,Williams2000,Abello1998,Broder2000,Urban2001}.

Among the various characteristic features that describe the behavior of random graphs, a large number of studies have been devoted to the analysis of its structural and dynamical properties~\cite{Strogatz2001,Dorogovtsev2001,Albert2002,Newman2003,Boccalettia2006,DurretBook,LatoraBook,EstradaBook,PenroseBook}. 
Also, some aspects of the transport properties of various random graph models have been considered since they provide further insight into the structure and function of the corresponding  graphs~\cite{Lopez2005,Lopez2006,Wu2006,Gallos2007,Candia2007,Ramasco2007,Li2007,Nicolaides2010,Mulken2011,Xiong2018}. 
In addition, the spectral and eigenfunction statistical properties of random graphs have also been analyzed~\cite{Zhu2000,Giraud2005,Sade2005,Slanina2012,Bandyopadhyay2007,Jalan2007,Jalan2009,deCarvalho2009,Jalan2010,Zhu2008,Jahnke2008,Goh2001,Rodgers2005,Nagao2008,Georgeot2010,Jalan2011,Gong2006,Farkas2002,Dorogovtsev2003,Martinez2019,Martinez2019}, in some cases with the help of random-matrix theory (RMT) measures and techniques~\cite{Zhu2000,Giraud2005,Sade2005,Slanina2012,Bandyopadhyay2007,Jalan2007,Jalan2009,deCarvalho2009,Jalan2010,Zhu2008,Jahnke2008,Martinez2019,Martinez2022}. 
The use of RMT has mainly been motivated by the equivalence between the adjacency matrix representing a given graph and the Hamiltonian~\cite{Rodgers1988,Fyodorov1991a,Fyodorov1991b,Mirlin1991,Evangelou1992a,Evangelou1992b,Jackson2001} representing the corresponding tight-binding structure. Here, both the adjacency matrix and the tight-binding Hamiltonian matrix (describing electronic systems found in condensed matter physics), are usually sparse random matrices.

Within the random graphs of current interest, much attention has been paid to the realistic situation in which the graphs are constrained by geometrical spatial bounds where, in addition, the connections between their constituents are determined by a distance threshold~\cite{Barthelemy2011,PenroseBook}. This family of graphs, that can be modeled by the so-called random geometric graph model~\cite{Gilbert1959} or variations of it~\cite{Barthelemy2011,PenroseBook,Dall2002,Stegehuis2022,Allen2018,Estrada2015,Pratt2018,Estrada2017,Waxman1988}, has found applications in a broad variety of scenarios. Even though, some spectral and eigenfunction properties of random geometric graphs have been studied from an RMT point of view, see for instance Refs.~\cite{Martinez2019,Alonso2018,Aguilar2020,PM23}, the electronic transport through the corresponding tight-binding structures has been left unexplored. Therefore, to take a step forward in understanding tight-binding random structures embedded in the plane, the purpose of the present paper is to analyze, by extensive numerical simulations, the scattering and transport statistical properties of random geometric graphs (RGGs) and bipartite random geometric graphs (BRGGs). The BRGG model is a variation of the RGG model, introduced to describe 5G wireless networks with multi-connectivity~\cite{Stegehuis2022}, containing vertices of two different species where connections are allowed between vertices of different species only.

Here, for the statistical analysis of scattering and transport properties of tight-binding random graphs, the RMT of electronic transport is used from which the main quantity of interest is the scattering matrix or $S$-matrix~\cite{Beenakker1997,MelloBook}. Indeed, all quantities discussed in this work are obtained from the $S$-matrix. In particular, the focus is paid on the average of the elements of the $S$-matrix, the averaged transmission, the channel-to-channel transmission distributions including the total transmission distribution, the shot-noise power, and the elastic enhancement factor. All these quantities will be analyzed as a function of the graph parameters and in a scattering setup of few open channels.

The organization of the paper is as follows. In the next Section, the random graph models as well as the scattering setup are described. A review of the analytical RMT predictions of the scattering and transport properties relevant to this study is provided in Sec.~\ref{sec:RMT}. The analysis of the scattering and transport properties of both graph models, RGGs and BRGGs, is the subject of Secs.~\ref{sec:StatisticsRGGs} and~\ref{sec:StatisticsBRGGs}, respectively. There, a smooth crossover from insulating to metallic behavior in the scattering and transport properties of both models is first observed. Then, once the graph properties are properly scaled, a universal insulating to metallic crossover is demonstrated. Finally, the conclusions are presented in Sec.~\ref{sec:Conclusions}.

It is important to add that a study of tight-binding Erd\"os-Renyi random graphs, similar to the one performed below for random geometric graphs, was reported in Ref.~\cite{Martinez2013}. There, it was shown that the average degree serves as the scaling parameter of scattering and transport properties of Erd\"os-Renyi graphs. However, at the end of this work, in the Conclusions section, that result will be further discussed since we believe it can be improved.


\section{Graph models and scattering setup}
\label{sec:Models}

As mentioned above, two graph models are considered in this work: RGGs and BRGGs.
On the one hand, RGGs consist of $N$ vertices uniformly and independently distributed in the unit square where two vertices are connected by an edge if their Euclidean distance is less or equal than the connection radius $r$~\cite{PenroseBook,Dall2002}. Thus, RGGs depend on the parameter pair $(N,r)$. On the other hand, BRGGs 
are composed by $N$ vertices grouped in two disjoint sets with $N-s$ and $s$ vertices each such that there are no adjacent vertices within the same set. The vertices belonging to both sets are uniformly and independently distributed in the unit square. Then, two vertices are connected by an edge if their Euclidean distance is less or equal than the connection radius $r$~\cite{Stegehuis2022}. Thus, BRGGs depend on the parameters $(N,s,r)$.

Here, within an RMT approach, the randomly weighted versions of RGGs and BRGGs are adopted for which the adjacency matrix of each graph model can be described by the following sparse tight-binding Hamiltonian
\begin{equation}
H = \sum_{n=1}^{N} h_{nn} | n \rangle \langle n | + \sum_{n, m=1}^{N} h_{nm} ( | n \rangle \langle m | + | m \rangle \langle n | ) ,
\label{eq:H}
\end{equation}
where $N$ is the number of vertices in the graph, $h_{nn}$ are on-site potentials, and $h_{nm}$ are hopping integrals between sites $n$ and $m$. Besides the fact that the vertices in each graph have random positions within the unit square, the weights $h_{nm}$ in Eq.~(\ref{eq:H}) are chosen to be statistically independent random variables drawn from a normal distribution with zero mean $\langle h_{nn} \rangle = 0$ and variance $\langle |h_{nm}|^{2} \rangle = (1 + \delta_{nm})/2$. This may correspond for instance to random self-loops and random strength interactions between vertices. Also, each graph is assumed to be undirected i.e., $h_{nm} = h_{mn}$ such that $H=H^{T}$, with $T$ the transpose operation, which in turn corresponds to time-reversal symmetric graphs. With this prescription (random self-loops $h_{nn}$ and random weight connections between vertices) for the model of Eq.~(\ref{eq:H}), well known RMT predictions for electronic transport are retrieved in the appropriate graph limits. Those predictions, discussed below, will be used as a reference throughout the paper.

The isolated graphs, represented by the tight-binding Hamiltonian of Eq.~(\ref{eq:H}), are open by attaching $2M$ semi-infinite single-channel leads to them; each lead is described by the one-dimensional tight-binding Hamiltonian
\begin{equation}
H_{\mathrm{lead}} = \sum_{n=1}^{-\infty} (| n \rangle \langle n + 1 | + | n + 1 \rangle \langle n |) .
\end{equation}
Then, the $2M \times 2M$ scattering matrix, $S(E)$, can be written as~\cite{Verbaarschot1985,Beenakker1997}
\begin{equation}
S(E) = 
\left(
\begin{array}{ccc}
r & t' \\
t & r'
\end{array}
\right) =
\openone_{2M} - 2\mathrm{i} \sin(k) \pi W^{T} \frac{1}{E - \mathcal{H}_{\mathrm{eff}} } W ,
\label{eq:S}
\end{equation}
where $r(r')$ and $t(t')$ are, respectively, $M\times M$ reflection and transmission matrices when the incidence is from the left (right) of the scattering region, $\openone_{2M}$ is the unit matrix of dimension $2M$, $k = \arccos(E/2)$ is the wave vector supported in the leads, $E$ is the energy, and $\mathcal{H}_\mathrm{eff}$ is the non-Hermitian effective Hamiltonian given by
\begin{equation}
H_\mathrm{eff} = H - \mathrm{i} \pi W W^{T} .
\label{eq:Heff}
\end{equation}
Here, $H$ is the $N\times N$ tight-binding Hamiltonian matrix of Eq.~(\ref{eq:H}) that describes the isolated graphs with $N$ resonant states. $W$ is an $N\times 2M$ energy independent matrix that couples those resonant states to the $M$ propagating modes in the leads. The elements of $W$ are $W_{ij} = \varepsilon \delta_{i,j_{0}}$, where $\varepsilon$ is the coupling strength between the graph and the leads and $j_{0} = 1, 2, \ldots , M$ specifies the vertices at which the leads are attached. Furthermore, assuming that the wave vector $k$ changes slightly in the center of the band, we set $E = 0$ and neglect the energy dependence of $H_{\mathrm{eff}}$ and $S$. For RGGs all vertices are equivalent, then the $2M$ leads are attached to $2M$ vertices chosen at random. However, since for BRGGs the vertices belong to two different disjoint sets, i.e.~vertices are not equivalent, then there are several ways to attach the $2M$ leads. For example, all leads could be attached to vertices of the larger set, all leads could be attached to vertices of the smaller set, or the leads could be attached to vertices of both sets in equal or different proportions. The particular scattering setup used in this work for BRGGs will be defined in Sec.~\ref{sec:StatisticsBRGGs}.

Moreover, since for RGGs and BRGGs characterized by a fixed parameter set, $(N,r)$ and $(N,s,r)$ respectively, the scattering and transport quantities fluctuate not only due to the random configurations of the vertices composing the graphs but also due to the random strength interactions between vertices, then for its analysis a statistical ensemble of equivalent graphs has to be considered. The fluctuations, that only depend on the time-reversal symmetry present in the graphs, can be characterized by the RMT of electronic transport for the appropriate symmetry class which in this case corresponds to the so-called Circular Orthogonal Ensemble (COE)~\cite{Beenakker1997}. Therefore, in what follows, a brief summary of the analytical predictions from the RMT of electronic transport for the COE is presented. We will see later that those predictions describe the scattering and transport properties of mostly connected graphs.


\section{RMT predictions of electronic transport for the COE}
\label{sec:RMT}

In this section we review known analytical predictions from the RMT of electronic transport through complex media in the presence of time-reversal symmetry, within the perfect coupling condition, $\langle S \rangle \approx 0$, between the scattering region (here the graph) and the outside (i.e.~the leads). In particular, the focus is paid on the average of the elements of the $S$-matrix, the channel-to-channel or mode-to-mode transmission distributions including the total transmission distribution, the averaged transmission, the shot-noise power, and the elastic enhancement factor.

The first quantity of interest is the average of the magnitude of the elements of the $S$-matrix, namely~\cite{Beenakker1997}
\begin{equation}
\langle |S_{mn}|^{2} \rangle_{\mathrm{COE}} = \frac{1 + \delta_{mn}}{2M + 1} ,
\label{eq:Smn}
\end{equation}
where $\langle \cdot \rangle$ denotes an ensemble average over the COE, $2M$ is the number of open channels, and $\delta_{mn}$ is the usual Kronecker delta. From the elements of the $S$-matrix, see Eq.~(\ref{eq:S}),  the transmission coefficient, or dimensionless conductance, can be obtained as~\cite{Beenakker1997,MelloBook}
\begin{equation}
T = \mathrm{Tr} (t t^{\dagger}) ,
\end{equation}
where Tr is the trace operation, $t$ is the transmission amplitude, and the symbol $\dagger$ represents the adjoint of $t$. The transmission distribution $w(T)$ for $M = 1$, i.e.~when two single-channel leads are attached to the complex scattering media, is given by~\cite{Beenakker1997,MelloBook}
\begin{equation}
w_{\mathrm{COE}}(T) = \frac{1}{2\sqrt{T}} , \quad \mathrm{with} \quad 0<T<1.
\label{eq:wTN1}
\end{equation}

Furthermore, in the $M=2$ case the transmission amplitude $t$ becomes a $2\times 2$ matrix and the following transmissions can be computed 
\begin{equation}
T_{nm} = |t_{nm}|^{2} , \,\, T_{n} = \sum_{m=1}^{2M} |t_{nm}|^{2}, \,\, \mathrm{and} \,\, T = \sum_{n,m}^{2M} |t_{nm}|^{2}  ,
\label{eq:speckle}
\end{equation}
which measure respectively the transmitted wave from mode $n$ to mode $m$, the transmitted wave from mode $n$ to mode $m$ if the incident wave is entirely in mode $n$, and the total transmission~\cite{vanLangen1996}. The respective probability distributions are given by~\cite{Pereyra1983}
\begin{eqnarray}
\label{eq:wTnm}
w_{\mathrm{COE}}(T_{nm}) & = & 6 (1 - \sqrt{T_{nm}})^{2} , \\
\label{eq:wTn}
w_{\mathrm{COE}}(T_{n}) & = & 6 \sqrt{T_{n}}  (1 - \sqrt{T_{nm}}), \quad \mathrm{and} \\
\label{eq:wTN2}
w_{\mathrm{COE}}(T) & = &
\begin{cases}
\frac{3}{2} T & \text{if } 0 < T < 1, \\
\frac{3}{2}( T - 2\sqrt{T-1} ) & \text{if } 1 < T < 2.
\end{cases}
\end{eqnarray}
In addition, for an arbitrary number of open channels, the average transmission is~\cite{MelloBook}
\begin{equation}
\langle T \rangle_{\mathrm{COE}} = \frac{M}{2} - \frac{M}{2(2M +1)} .
\label{eq:AvgT}
\end{equation}

Another transport quantity of interest is the so-called shot noise power $P$, defined as $P = \langle \mathrm{Tr} (t t^{\dagger} - t t^{\dagger} t t^{\dagger} ) \rangle$, 
which as a function of $M$ reads~\cite{Savin2006}
\begin{equation}
P_{\mathrm{COE}} = \frac{M (M + 1)^{2}}{2 (2M + 1) (2M + 3)} .
\label{eq:P}
\end{equation}
Finally, a measure of the fluctuations of the scattering cross-section is provided by the elastic enhancement factor defined as
\begin{equation}
F = \frac{\langle |S_{mm}|^{2} \rangle}{\langle |S_{mn}|^{2} \rangle } ,
\end{equation}
which in the RMT limit becomes~\cite{Harney1986}
\begin{equation}
F_{\mathrm{COE}} = 2 .
\label{eq:F}
\end{equation}

In the following sections, the scattering and transport properties of tight-binding random graphs, described by Eq.~(\ref{eq:H}) for both RGG and RBGG models, will be computed. The RMT predictions given in Eqs.~(\ref{eq:Smn}-\ref{eq:F}) are expected to be retrieved for mostly connected graphs in the perfect coupling regime characterized by $\langle S \rangle \approx 0$.


\section{Scattering and transport properties of tight-binding RGGs}
\label{sec:StatisticsRGGs}

For the statistical analysis shown throughout this work, the calculations are performed at $E = 0$ considering graph sizes of $N = 100$, 200, and 400, with random realizations or ensemble sizes of $10^{4}$, $5\times 10^{3}$, and $2.5\times 10^{3}$, respectively. Also, the results shown below do not contain error bars since the statistics is done with a large amount of data such that the error is not significant.


\subsection{Perfect coupling condition for RGGs}

%
\begin{figure}
\centering
\includegraphics[width=0.6\columnwidth]{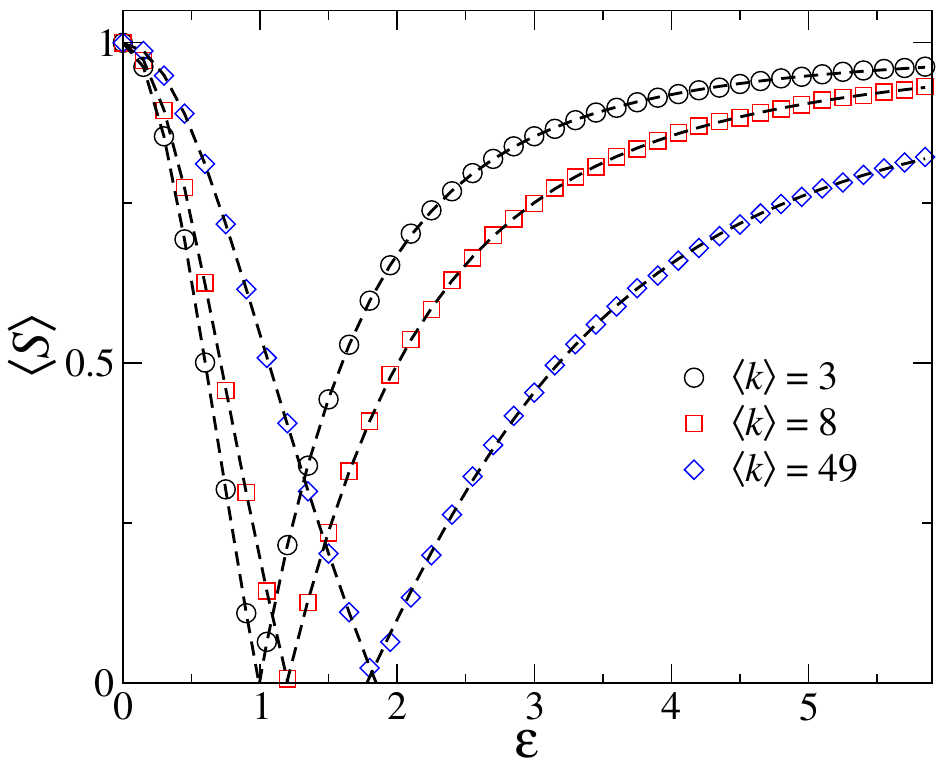}
\caption{Average $S$-matrix as a function of the coupling strength $\varepsilon$ for tight-binding RGGs of size $N=100$. The scattering setup with two single-channel leads ($M=1$) is considered. Three values of the average degree are reported: $\langle k \rangle = 3$, 8, and 49. The black-dashed curves are fittings of Eq.~(\ref{eq:FittingS}) to the numerical data (symbols). The error bars are smaller than the symbol size, so they are not displayed. The coupling strength $\varepsilon = \varepsilon_{0}$ such that $\langle S \rangle \approx 0$ sets the perfect coupling condition between the graphs and the leads.}
\label{fig:RGGsAvgS}
\end{figure}

Before proceeding with the statistical analysis of the scattering and transport properties of RGGs, in order to be able to compare the numerical results with the RMT predictions of the preceding section, the perfect coupling condition is first established. This is performed by setting the coupling strength $\varepsilon = \varepsilon_{0}$ such that the average of the scattering matrix, $\langle S \rangle$ (also known as optical matrix), vanishes. That is
\begin{equation}
\langle S \rangle \equiv \frac{1}{2M} \sum_{m, n} |\langle S_{mn} \rangle| \approx 0 ,
\label{eq:PerfectS}
\end{equation}
where the average is taken over the whole ensemble of random graphs. 

Then, in Fig.~\ref{fig:RGGsAvgS} the average of the $S$-matrix for tight-binding RGGs as a function of the coupling strength $\varepsilon$ is shown. Here, the $M=1$ case has been considered, i.e.~when two single-channel leads are attached to the graphs. Without loss of generality RGGs with $N=100$ vertices and average degrees $\langle k \rangle = 3$ (black circles), 8 (red squares), and 49 (blue rhomboids) were used. As expected, for the coupling strength $\varepsilon = 0$ the graphs and the leads are disconnected and thus $\langle S \rangle = 1$. That is, for $\varepsilon = 0$ the waves in the leads are fully reflected back before they interact with the graphs. As the coupling strength $\varepsilon$ increases, $\langle S \rangle$ decays until it reaches its minimum value $\langle S \rangle \approx 0$ when $\varepsilon = \varepsilon_{0}$. This coupling strength defines the perfect coupling condition between the leads and the graphs. Indeed, at $\varepsilon = \varepsilon_{0}$, the waves in the leads fully enter the scattering region and interact with it. For values of $\varepsilon > \varepsilon_{0}$, $\langle S \rangle$ increases. As observed in Fig.~\ref{fig:RGGsAvgS}, the curves of $\langle S \rangle$ vs.~$\varepsilon$ for different $\langle k \rangle$ show a similar behavior which is well fitted by the following heuristic expression~\cite{Martinez2013}
\begin{equation}
\langle S \rangle = \frac{c_{0}}{1 + (c_{1} \varepsilon)^{\pm c_{2}} } - c_{3} ,
\label{eq:FittingS}
\end{equation}
where the $c_{i}$ are fitting constants and the plus (minus) sign corresponds to the region $0 < \varepsilon < \varepsilon_{0}$ ($\varepsilon > \varepsilon_{0}$). As in~\cite{Martinez2013}, here we use Eq.~(\ref{eq:FittingS}) to extract $\varepsilon_{0}$ from a small set of data values $\langle S \rangle$ vs.~$\varepsilon$. As examples, in Fig.~\ref{fig:RGGsAvgS} the dashed lines correspond to fittings of~(\ref{eq:FittingS}) to the numerical data (symbols). For the RGGs with average degrees $\langle k \rangle = 3$, 8, and 49, the fitting constants ($c_{0}, c_{1}, c_{2}, c_{3}$) are respectively ($2.29, 0.88, 2.03, 1.29$), ($1.70, 0.94, 2.22, 0.71$), and ($1.99, 0.54, 2.00, 0.99$). For these cases, the obtained coupling strengths, such that $\langle S \rangle \approx 0$, are respectively $\varepsilon_{0} = 0.9893$, 1.2371, and 1.840. Furthermore, from a detailed numerical analysis (not shown here) of $\varepsilon_{0}$ vs.~$\langle k \rangle$, we found that
\begin{equation}
\varepsilon_{0} \approx 0.277 \langle k \rangle^{2/5} + 0.598 ,
\label{eq:Epsilon0}
\end{equation}
works well to compute $\varepsilon_{0}$ given the average degree $\langle k \rangle$ irrespective of the graph size and the number of attached leads. Thus, in what follows, we use Eq.~(\ref{eq:Epsilon0}) to set the perfect coupling condition of the scattering setup.

%
\begin{figure}
\centering
\includegraphics[width=0.98\columnwidth]{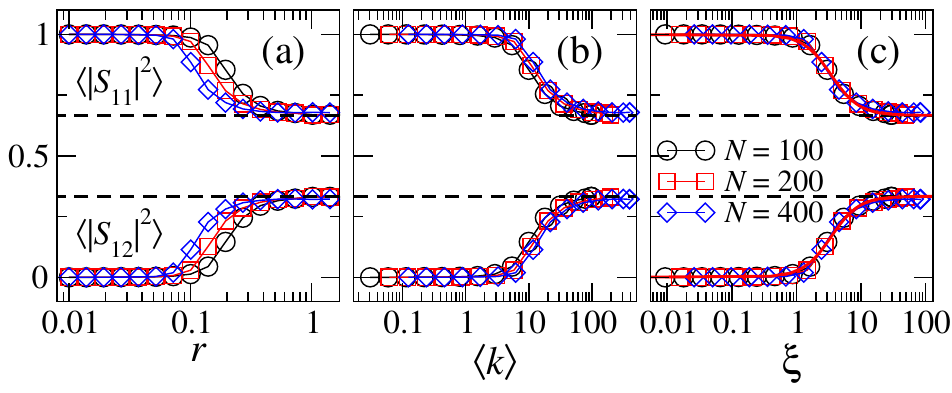}
\caption{(Color online) Absolute value of the averaged $S$-matrix elements $\langle |S_{11}|^{2} \rangle$ and $\langle |S_{12}|^{2} \rangle$ for tight-binding RGGs as a function of (a) the connection radius $r$, (b) the average degree $\langle k \rangle$, and (c) the scaling parameter $\xi$. Three graphs sizes $N$ are reported: $N=100$, 200, and 400. The scattering setup with two single-channel leads ($M=1$) is considered. The dashed lines correspond to the respective RMT predictions given by Eq.~(\ref{eq:Smn}). The red continuous lines in panel (c) are fittings of Eq.~(\ref{eq:relSmnM}) to the data with fitting parameter $\delta = 0.29$ and statistical indicator of $\chi^{2} = 4\times 10^{-4}$. In all cases, a smooth transition from an insulating to a metallic behavior, from nearly disconnected to mostly connected graphs, is observed.}
\label{fig:RGGsAvgSmnM1}
\end{figure}
%


\subsection{Scattering elements of RGGs and scaling parameter}

In Fig.~\ref{fig:RGGsAvgSmnM1} the averaged squared modulus of the $S$-matrix elements, $\langle |S_{11}|^{2} \rangle$ and $\langle |S_{12}|^{2} \rangle$, for RGGs  is shown as a function of the connection radius $r$ (panel (a)) and the average degree $\langle k \rangle$ (panel (b)). Since the scattering setup with two single-channel leads ($M=1$) is considered, the elements $\langle |S_{11}|^{2} \rangle$ and $\langle |S_{12}|^{2} \rangle$ correspond to the averaged reflection and averaged transmission coefficients respectively. For nearly disconnected graphs ($r \ll 1$ or $\langle k \rangle \ll 1$), the waves in the leads are not transmitted and thus $\langle |S_{12}|^{2} \rangle \approx 0$. As the graphs become more connected, the transmission $\langle |S_{12}|^{2} \rangle$ increases until it reaches the maximum value of 1/3, expected for complete graphs, in agreement with RMT predictions; see Eq.~(\ref{eq:Smn}). For the averaged reflection coefficient, the opposite behavior is observed: For nearly disconnected graphs $\langle |S_{11}|^{2} \rangle \approx 1$ since the waves in the leads are reflected back before entering the scattering region. As the connectivity of the graphs increases, the reflection coefficient decreases until it reaches the value of 2/3 predicted by RMT; see Eq.~(\ref{eq:Smn}). That is, $\langle |S_{11}|^{2} \rangle$ and $\langle |S_{12}|^{2} \rangle$ show a smooth transition from an insulating to a metallic behavior as a function of the connection radius $r$, or the average connectivity $\langle k \rangle$, from nearly disconnected to mostly connected graphs.

%
\begin{figure}
\centering
\includegraphics[width=1.0\columnwidth]{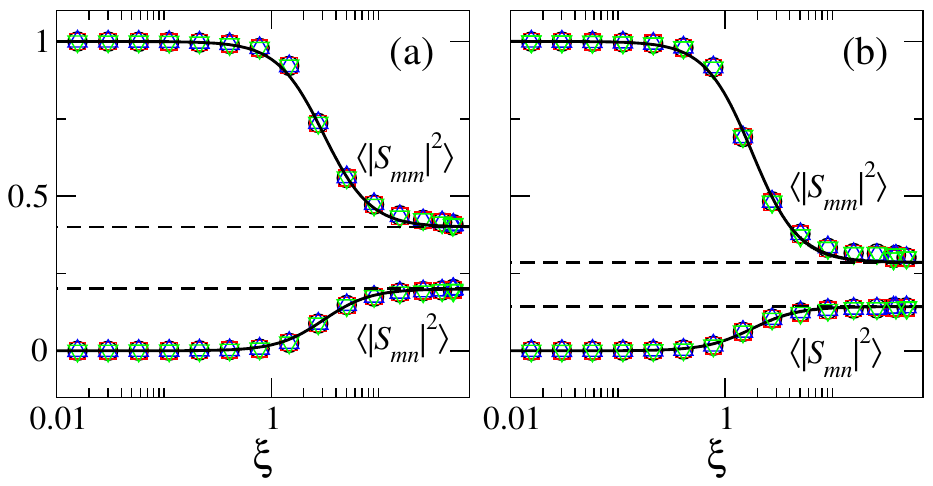}
\caption{(Color online) Average $S$-matrix elements $\langle |S_{mm}|^{2} \rangle$ and $\langle |S_{mn}|^{2} \rangle$ as a function of the scaling parameter $\xi$ for tight-binding RGGs with $N=200$ and (a) $M=2$ (four single-channel leads) and (b) $M=3$ (six single-channel leads). The scattering matrix elements are $\langle |S_{mm}|^{2} \rangle$ with $m=1$ (black circles), 2 (red squares), 3 (blue up-triangles) and 4 (green down-triangles); and $\langle |S_{mn}|^{2} \rangle$ with $mn = 12$ (black circles), 23 (red squares), 34 (blue up-triangles) and 41 (green down-triangles). The black-dashed lines are the corresponding RMT predictions from Eq.~(\ref{eq:Smn}). The black continuous lines are fittings of Eq.~(\ref{eq:relSmnM}) to the data with fitting constant $\delta = 0.325$ (0.562) and statistical indicator of $\chi^{2} = 5\times 10^{-5}$ ($= 2\times 10^{-4}$) for the $M=2$ (3) case, respectively. As the scaling parameter $\xi$ increases, a smooth transition from insulating to metallic behavior is observed in all cases, as is also observed in Fig.~\ref{fig:RGGsAvgSmnM1}.}
\label{fig:RGGsAvgSmnM2M3}
\end{figure}

Also, in Fig.~\ref{fig:RGGsAvgSmnM1} panels (a) and (b), $\langle |S_{11}|^{2} \rangle$ and $\langle |S_{12}|^{2} \rangle$ show a dependence on the graph size; as a function of $r$ the larger the graph size, the faster the RMT prediction is reached (the opposite behavior occurs as a function of $\langle k \rangle$), even though this dependence is barely visible as a function of $\langle k \rangle$. Moreover, note that the shape of the curves of $\langle |S_{11}|^{2} \rangle$ and $\langle |S_{12}|^{2} \rangle$ as a function of $r$ do not change despite the fact that $N$ varies, which means that these curves should be scale invariant. Indeed, for Erd\"os-R\'enyi graphs, it has been shown that the average degree $\langle k \rangle$ is the parameter that scales the scattering and transport properties~\cite{Martinez2013}. Here, however, $\langle k \rangle$ does not scale the $S$-matrix elements of  tight-binding RGGs as observed in Fig.~\ref{fig:RGGsAvgSmnM1}(b). Instead, the parameter
\begin{equation}
\xi \equiv \langle k \rangle N^{-\alpha},
\label{eq:xi}
\end{equation}
which depends on both the average degree $\langle k \rangle$ and the graph size $N$, properly scales the $S$-matrix elements with $\alpha = 0.26 \pm 0.0185$; as shown in Fig.~\ref{fig:RGGsAvgSmnM1}(c). Furthermore, this universal behavior of $\langle |S_{11}|^{2} \rangle$ and $\langle |S_{12}|^{2} \rangle$ as a function of $\xi$ is very well described by~\cite{Martinez2013}
\begin{eqnarray}
\langle |S_{mm}|^{2} \rangle & = & 1 - (2M - 1) \langle |S_{nm}|^{2} \rangle \quad \mathrm{and} \nonumber \\
\langle |S_{nm}|^{2} \rangle & = & \langle |S_{nm}|^{2} \rangle_{\mathrm{COE}} \frac{1}{1 + (\delta \xi)^{-2}} ,
\label{eq:relSmnM}
\end{eqnarray}
where $2M$ is the number of open channels, $\langle |S_{nm}|^{2} \rangle_{\mathrm{COE}}$ is the RMT prediction of $\langle |S_{nm}|^{2} \rangle$ for the COE given by Eq.~(\ref{eq:Smn}), and $\delta$ is a fitting parameter. Indeed, in Fig.~\ref{fig:RGGsAvgSmnM1}(c), relations~(\ref{eq:relSmnM}) for $\langle |S_{11}|^{2} \rangle$ and $\langle |S_{12}|^{2} \rangle$ with $\delta = 0.29$ are shown in red continuous lines. A good correspondence between~(\ref{eq:relSmnM}) and the numerics is observed. It is worth mentioning that expressions~(\ref{eq:relSmnM}) also describe well the average $S$-matrix elements of other random matrix models showing a metal-insulator transition~\cite{Martinez2013,Mendez2010,AMMA2023}.

%
\begin{figure}
\centering
\includegraphics[width=1.0\columnwidth]{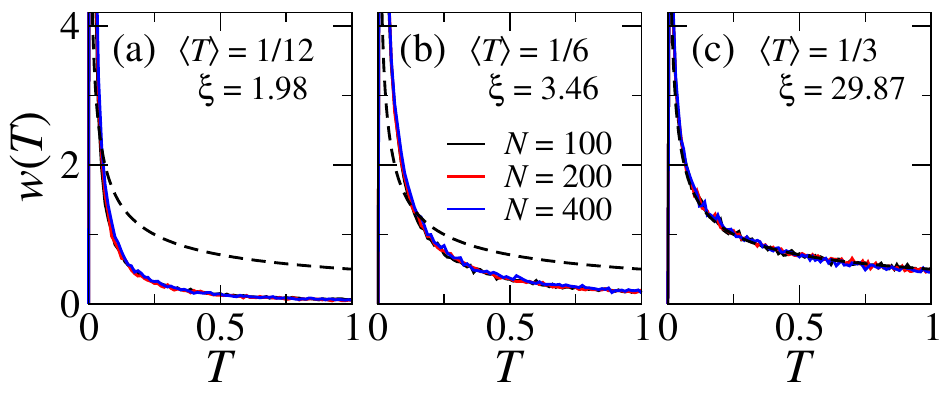}
\caption{(Color online) Transmission distribution $w(T)$ for tight-binding RGGs in the two open-channels setup ($M=1$). The numerical results correspond to graph sizes of $N = 100$, 200, and 400, depicted in black, red, and blue histograms. Three fixed values of the average transmission are considered: (a) $\langle T \rangle = $ 1/12, (b) 1/6, and (c) 1/3. These values yield respectively to nearly disconnected, half connected, and mostly connected graphs. The black dashed lines correspond to the RMT prediction of Eq.~(\ref{eq:wTN1}). The RMT limit is recovered for mostly connected graphs, see panel (c).}
\label{fig:RGGswTM1}
\end{figure}

In addition, in Fig.~\ref{fig:RGGsAvgSmnM2M3}, the scattering matrix elements $\langle |S_{mm}|^{2} \rangle$ and $\langle |S_{mn}|^{2} \rangle$ for tight-binding RGGs as a function of the scaling parameter $\xi$ are shown. Here, RGGs with $N=200$ vertices in a scattering setup with (a) four single-channel leads ($M=2$) and (b) six single-channel leads ($M=3$) are considered. A smooth transition from insulating to metallic behavior of both $\langle |S_{mm}|^{2} \rangle$ and $\langle |S_{mn}|^{2} \rangle$ is observed as the scaling parameter $\xi$ increases; similar to what happens for the $M=1$ case reported in Fig.~\ref{fig:RGGsAvgSmnM1}. Again, a good agreement between the numerical results and the RMT prediction given by Eq.~(\ref{eq:Smn}) is obtained for mostly connected graphs, specifically when $\xi \geq 10$.


\subsection{Transmission and shot noise power for RGGs}

Once we have stated that the parameter $\xi$ scales the averaged scattering matrix elements, we anticipate
that other related quantities will also scale with $\xi$.

%
\begin{figure}
\centering
\includegraphics[width=1.0\columnwidth]{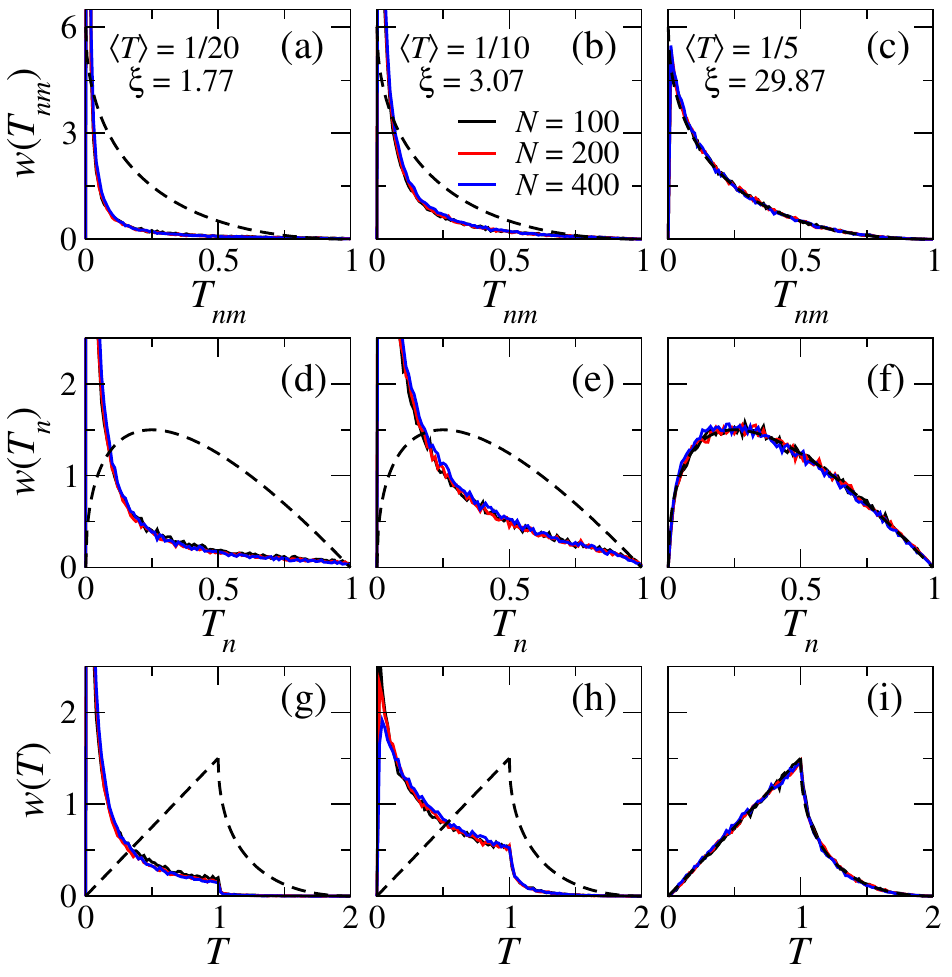}
\caption{(Color online) Channel-to-channel or mode-to-mode transmission distributions for tight-binding RGGs with $M=2$ (four open-channels). The numerical results correspond to RGGs of sizes $N = 100$, 200 and 400, shown in black, red and blue histograms, respectively. In the left, the middle, and the right panels the average transmission is fixed to $\langle T \rangle = $ 1/20, 1/10, and 1/5 which yields to nearly disconnected, half connected, and mostly connected graphs. Those values of $\langle T \rangle$ lead to $\xi = 1.77$, 3.07, and 29.87 respectively. The black dashed lines correspond to the respective RMT predictions given by Eqs.~(\ref{eq:wTn}-\ref{eq:wTN2}). The RMT limit is recovered for mostly connected graphs, see panels (c), (f), and (i).}
\label{fig:RGGswTM2}
\end{figure}

Then, in Fig.~\ref{fig:RGGswTM1}, the transmission distribution for tight-binding RGGs with two single-channel leads attached ($M=1$) is shown. Three fixed values of the average transmission are considered: $\langle T \rangle = 1/12$, 1/6, and 1/3, which yield to $\xi = 1.98$, 3.46, and 29.87, respectively. These values of $\xi$ are obtained from Eq.~(\ref{eq:relSmnM}) with $\delta = 0.29$, taken from the fitting of Eq.~(\ref{eq:relSmnM}) to the numerical data for the scattering element $\langle |S_{12}| \rangle=\langle T \rangle$ (see Fig.~\ref{fig:RGGsAvgSmnM1}). It is worth stressing that once $\xi$ is fixed, the transmission distribution $w(T)$ becomes invariant; i.e., independent of the graph size. Notice that in each panel of Fig.~\ref{fig:RGGswTM1} three histograms (characterized by $N = 100$, 200, and 400) are reported, however they fall one on top of the other. Also, the RMT limit is reached for mostly connected graphs as observed in Fig.~\ref{fig:RGGswTM1}(c).

%
\begin{figure}
\centering
\includegraphics[width=1.0\columnwidth]{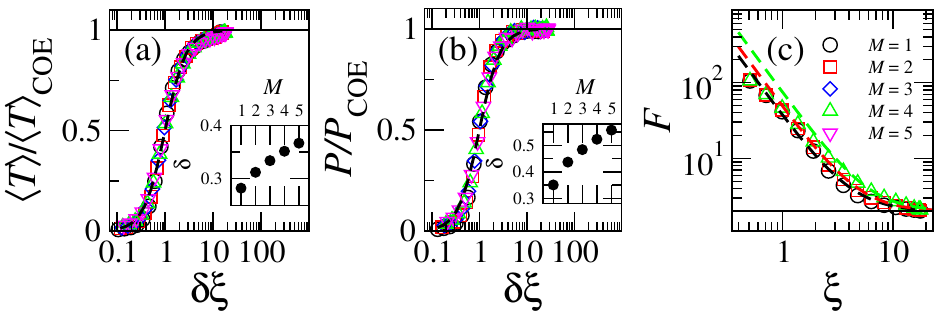}
\caption{(Color online) (a) Average transmission $\langle T \rangle$ and (b) shot noise power $P$ (normalized to their respective RMT limit values given by Eqs.~(\ref{eq:AvgT}) and (\ref{eq:P}), respectively) as a function of $\delta \xi$ for tight-binding RGGs with $N = 200$. The elastic enhancement factor $F$ as a function of $\xi$ is reported in (c). In (a) and (b) $2M$ single-channel leads are attached to the graphs with $M = 1, \ldots, 5$. The black dashed lines correspond to the fitting of Eq.~(\ref{eq:Xxi}) to the numerical data. The value of the fitting parameter $\delta$ for each case is shown in the corresponding insets. The black full lines are shown to guide the eye. Without loss of generality, the elastic enhancement factor in (c) is reported for $M = 1$, 2, and 4 only. The black, red, and green dashed lines correspond to fittings of Eq.~(\ref{eq:FitF}) to the numerical data with $c = 50.96, 47.82$, and 50.70, respectively. In all these cases the RMT limit value of $F = 2$ is reached for large $\xi$.}
\label{fig:RGGsTPF}
\end{figure}

For the $M = 2$ case, or for graphs with four single-channel leads attached to them, three types of transmissions can be computed whose respective distributions are given by Eqs.~(\ref{eq:wTn}-\ref{eq:wTN2}). Those mode-to-mode, or lead-to-lead, transmission distributions are shown in Fig.~\ref{fig:RGGswTM2} for RGGs of sizes $N = 100$, 200, and 400 in black, red, and blue histograms, compared to their corresponding RMT predictions given by Eqs.~(\ref{eq:wTn}-\ref{eq:wTN2}) in black dashed lines. For the left, middle, and right panels, the average transmission is fixed to $\langle T \rangle = 1/20$, 1/10, 1/5, which yield to the values of $\xi = 1.77$, 3.07, and 29.87, respectively. Again, once $\xi$, or equivalently $\langle T \rangle$, is fixed, the distributions become invariant or independent of the graph size. Furthermore, from the figure it is noticeable that as the graphs become more connected, or as $\xi$ increases, the divergency that occurs at zero transmission tends to disappear. Also, in all cases the numerical transmission distributions are well described by their corresponding RMT predictions for mostly connected graphs; see the right panels in Fig.~\ref{fig:RGGswTM2}.

Now, we turn our attention to the case of $M>2$. As reported in previous studies on the scattering and transport properties of tight-binding Erd\"os-R\'enyi random graphs~\cite{Martinez2013}, the average transmission and shot noise power follow a universal behavior as a function of $\delta \xi$ described by
\begin{equation}
X(\xi) = X_{\mathrm{COE}} \left[ \frac{1}{1 + (\delta \xi)^{-2}} \right] ,
\label{eq:Xxi}
\end{equation}
where $X$ represents either $\langle T \rangle$ or $P$, $\delta$ is a fitting parameter, and $X_{\mathrm{COE}}$ is the corresponding RMT limit value given by Eqs.~(\ref{eq:AvgT}) and (\ref{eq:P}), respectively.

The numerical results for tight-binding RGGs of size $N = 200$ are shown in Fig.~\ref{fig:RGGsTPF}(a) and Fig.~\ref{fig:RGGsTPF}(b) for the average transmission and shot noise power as a function of $\delta \xi$. Both quantities are normalized to their corresponding RMT limit values given in Eqs.~(\ref{eq:AvgT}) and (\ref{eq:P}), respectively. The number of channels supported by the graphs are $2M = 2, 4, \ldots, 10$ (symbols). As can be clearly observed in Fig.~\ref{fig:RGGsTPF}, both quantities $\langle T \rangle$ and $P$ do not depend on $M$ when scaled by $\delta \xi$ and behave similarly: They are approximately zero for nearly isolated graphs and grow until they reach the corresponding RMT values when the graphs are mostly connected. That is, a smooth transition from insulating to metallic behavior is also observed in those quantities. Furthermore, this behavior is very well described by expression~(\ref{eq:Xxi}) shown in black dashed lines in Figs.~\ref{fig:RGGsTPF}(a) and~\ref{fig:RGGsTPF}(b) which correspond to a fitting of~(\ref{eq:Xxi}) to the respective numerical data.

Finally, in Fig.~\ref{fig:RGGsTPF}(c) the elastic enhancement factor $F$, see Eq.~(\ref{eq:F}), as function of $\xi$ is shown. The symbols correspond to the numerical data for RGGs with $N = 200$ vertices supporting $2M = 2$, 4, and 8 open-channels. It is observed that as $\xi$ increases, $F$ decreases until it reaches the value of 2 for mostly connected graphs as predicted by RMT, see Eq.~(\ref{eq:F}).

Notice that by substituting Eqs.~(\ref{eq:relSmnM}) into Eq.~(\ref{eq:F}) one gets
\begin{equation}
F \approx (2M + 1) (\delta \xi)^{-2} + 2 ,
\end{equation}
which captures well the behavior of $F$ only for small and large $\xi$ values, but yields to discrepancies for intermediate $\xi$ (not shown here). As pointed out in Ref.~\cite{Martinez2013}, this discrepancy comes from an overestimation of $\langle |S_{mn}|^{2} \rangle$ for $\xi \sim 1$. A better heuristic estimation of the behavior of $F$ was proposed to be~\cite{Martinez2013}
\begin{equation}
F \approx c\, \xi^{-2} + 2 ,
\label{eq:FitF}
\end{equation}
with $c$ a fitting constant. In Fig.~\ref{fig:RGGsTPF}(c), the fitting of the numerical data to expression~(\ref{eq:FitF}) is shown for $M = 1,$ 2, and 4, in black, red, and green dashed lines where $c \approx 50$. As observed, a reasonably good description of the behavior of $F$ is provided by~(\ref{eq:FitF}).

%
\begin{figure}
\centering
\includegraphics[width=1.0\columnwidth]{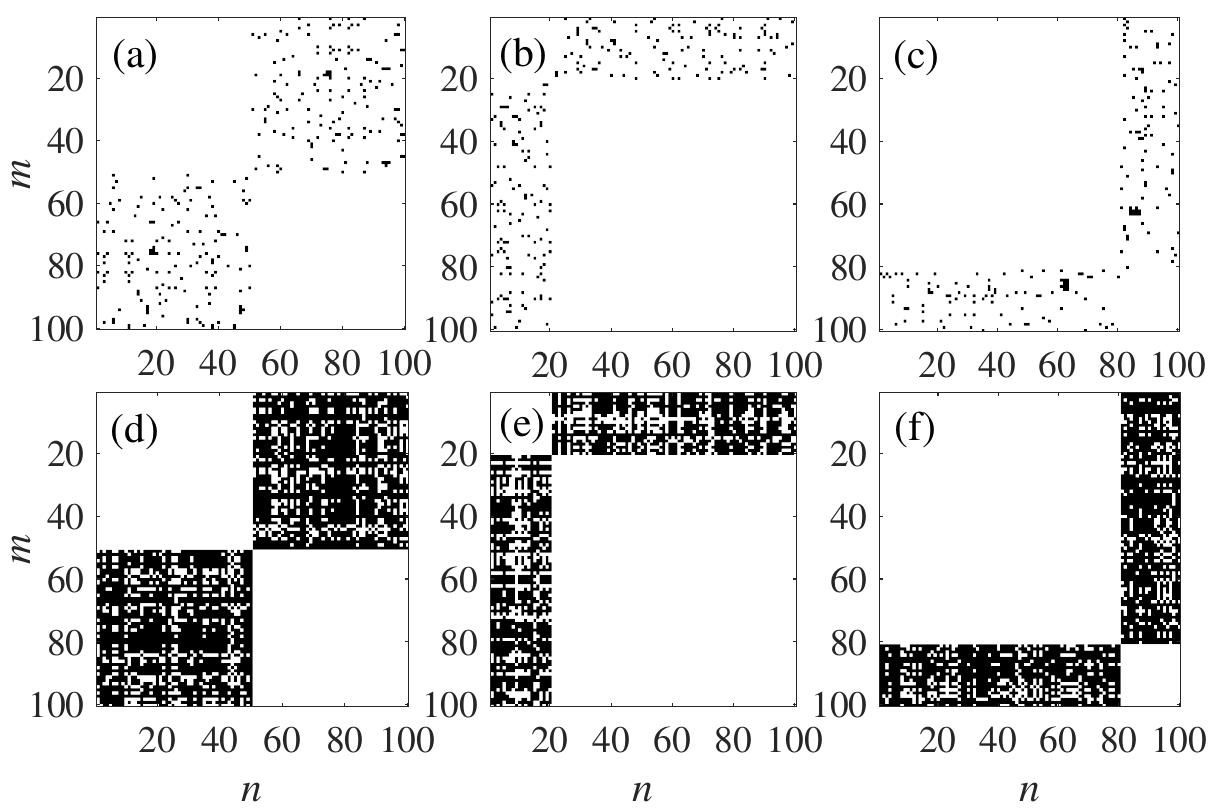}
\caption{Nonzero adjacency matrix elements of BRGGs of size $N = 100$ and (a, d) BRGG($N/2$), (b, e) BRGG($N/5$), and (c, f) BRGG($4N/5$). Two values of the connection radius are reported: (upper panels) $r = 0.15$ and (lower panels) $r = 0.7$. A single realization of BRGGs is used in each panel.}
\label{fig:NEWFIG}
\end{figure}
%


\section{Scattering and transport properties of tight-binding BRGGs}
\label{sec:StatisticsBRGGs}

As already defined in the Introduction, BRGGs are composed by $N$ vertices grouped in two disjoint sets: set A and set B having $s$ and $N-s$ vertices, respectively, such that there are no adjacent vertices within the same set. In this respect, the case $s = N/2$ is a limiting case where both sets have the same number of vertices. The vertices belonging to both sets are uniformly and independently distributed in the unit square where two vertices are connected by an edge if their Euclidean distance is less or equal than the connection radius $r$. Thus, BRGGs depend on the parameters $(N,s,r)$. BRGGs were named as AB random geometric graphs in Ref.~\cite{Stegehuis2022} to stress the fact that the vertices belong to two different sets, set A and set B.

In what follows, given BRGGs of total size $N$, three different sizes of the set A are considered: $s = N/2$, $N/5$, and $4N/5$. Hereafter, these BRGGs will be named as BRGG($s$). As examples, in Fig.~\ref{fig:NEWFIG} the adjacency matrices of BRGGs of size $N = 100$ and $s = N/2$, $N/5$, and $4N/5$ are shown for two values of the connection radius. Then, the $2M$ single-channel leads are attached to randomly selected vertices of set A; and the set B is of size $N-s$. Therefore, while for $s = N/2$ (when both sets have the same size) it is not relevant to which set the leads are attached, for $s=N/5$ ($s=4N/5$) the leads are attached to vertices of the smaller (larger) set.

As in the statistical analysis of RGGs, in order to compare the numerical results with the RMT predictions, the perfect coupling condition is first set. Also, for the statistical analysis the calculations are performed at $E = 0$ considering BRGGs of sizes $N = 100$, 200, and 400 with ensemble sizes of $10^{4}$, $5\times 10^{3}$, and $2.5 \times 10^{3}$ for each parameter combination. The results shown below do not contain error bars since the statistics is done with a large amount of data such that the error is not significant.

%
\begin{figure}
\centering
\includegraphics[width=1.0\columnwidth]{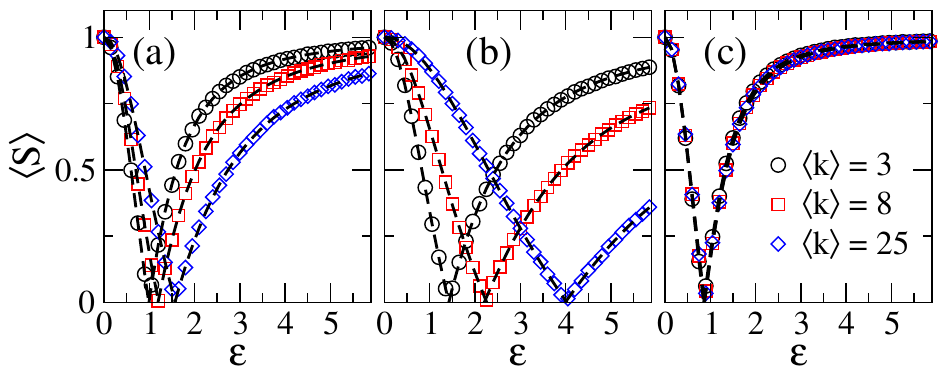}
\caption{(Color online) Average $S$-matrix as a function of the coupling strength $\varepsilon$ for tight-binding BRGGs of size $N = 100$ and (a) BRGG($N/2$), (b) BRGG($N/5$), and (c) BRGG($4N/5$). The scattering setup with two single-channel leads $(M = 1)$ is considered. Three values of the average degree are reported: $\langle k \rangle=3$, 8, and 25. The black-dashed curves are fittings of Eq.~(\ref{eq:FittingS}) to the respective numerical data (symbols). The error bars are smaller than the symbols size, so they are not displayed.}
\label{fig:BRGGsAvgS}
\end{figure}
%


\subsection{Perfect coupling condition for BRGGs}

%
\begin{figure}
\centering
\includegraphics[width=1.0\columnwidth]{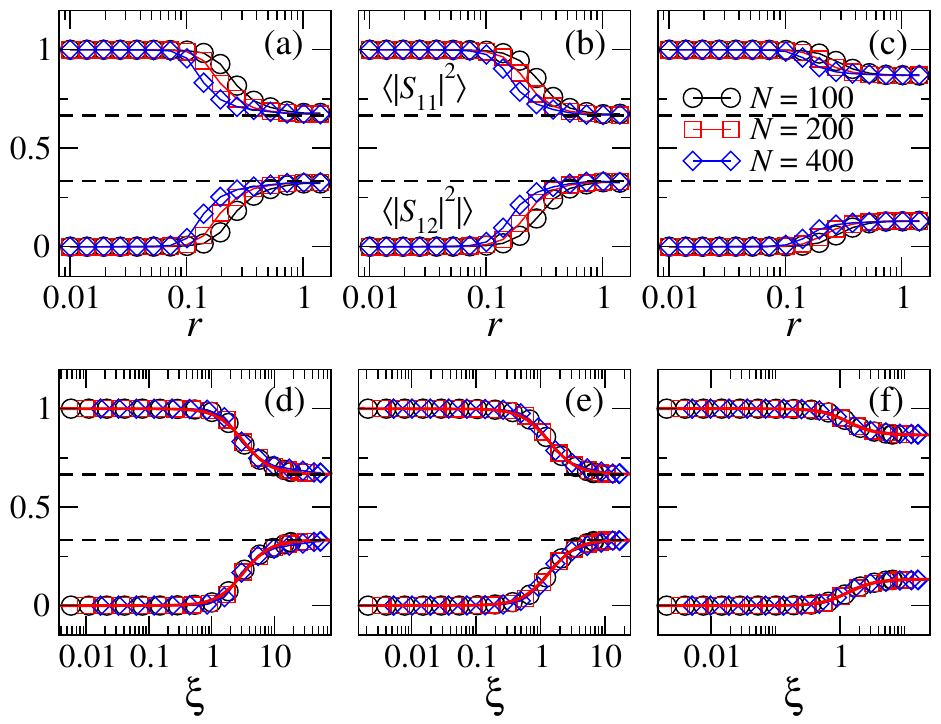}
\caption{(Color online) Absolute value of the averaged $S$-matrix elements $\langle |S_{11}|^{2} \rangle$ and $\langle |S_{12}|^{2} \rangle$ for tight-binding BRGGs as a function of (upper panels) the connection radius $r$ and (lower panels) the scaling parameter $\xi$. Three graphs sizes $N$ are reported ($N=100$, 200, and 400) for (left panels) BRGG($N/2$), (middle panels) BRGG($N/5$), and (right panels) BRGG($4N/5$). The scattering setup with two single-channel leads $(M = 1)$ is considered. The black dashed lines correspond to the respective RMT predictions given by Eq.~(\ref{eq:Smn}). The red continuous lines in the lower panels correspond to fittings of expression~(\ref{eq:relSmnM}) to the numerical data with fitting constants $\delta = 0.310$, 0.737, and 0.741 and statistical indicators of $\chi^{2} = 5\times 10^{-4}, 3\times 10^{-4},$ and $6\times 10^{-5}$ for BRGG($N/2$), BRGG($N/5$), and BRGG($4N/5$), respectively.}
\label{fig:BRGGsAvgSmnM1}
\end{figure}
%

%
\begin{figure}
\centering
\includegraphics[width=1.0\columnwidth]{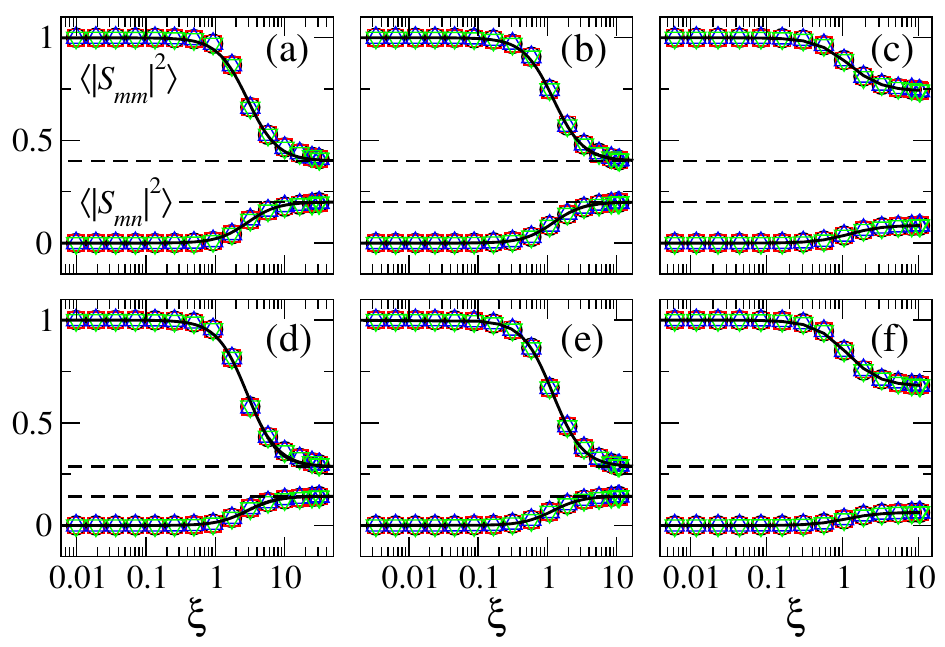}
\caption{(Color online) Average $S$-matrix elements $\langle |S_{mm}|^{2} \rangle$ and $\langle |S_{mn}|^{2} \rangle$ as a function of the scaling parameter $\xi$ for tight-binding BRGGs with $N=200$ and (upper panels) $M=2$ (four single-channel leads) and (lower panels) $M=3$ (six single-channel leads) open channels.
(Left panels) BRGG($N/2$), (middle panels) BRGG($N/5$), and (right panels) BRGG($4N/5$).
The scattering elements are $\langle |S_{mm}|^{2} \rangle$ with $m=1$ (black circles), 2 (red squares), 3 (blue up-triangles) and 4 (green down-triangles); and $\langle |S_{mn}|^{2} \rangle$ with $mn = 12$ (black circles), 23 (red squares), 34 (blue up-triangles) and 41 (green down-triangles). 
The black dashed lines correspond to the respective RMT predictions given by Eq.~(\ref{eq:Smn}). The black continuous lines are fittings of Eq.~(\ref{eq:relSmnM}) to the numerical data. The values of the fitting constants for the $M = 2$ (3) case are $\delta = 0.344$ (0.348), 0.818 (0.838), and 0.828 (0.880) with corresponding statistical indicators of $\chi^{2} = 1\times 10^{-4}\, (1\times 10^{-4}), 2\times 10^{-5} (1\times 10^{-5}),$ and $1\times 10^{-5}\, (1\times 10^{-5})$ for BRGG($N/2$), BRGG($N/5$), and BRGG($4N/5$), respectively.}
\label{fig:BRGGsAvgSmnM2M3}
\end{figure}

As for RGGs, in order to compare the numerical results from BRGGs with the analytical predictions from RMT (in the appropriate limits) discussed in Sect.~\ref{sec:RMT}, the perfect coupling condition between the BRGGs and the single-channel leads is first set. As before, this is achieved by finding the coupling strength $\varepsilon = \varepsilon_{0}$ such that $\langle S \rangle \approx 0$, see Eq.~(\ref{eq:PerfectS}).

Figure~\ref{fig:BRGGsAvgS} shows the average $S$-matrix as a function of the coupling strength $\varepsilon$ for tight-binding BRGGs in the three configurations to be explored: (a) BRGG($N/2$), (b) BRGG($N/5$), and (c) BRGG($4N/5$). The BRGGs have a total size of $N = 100$ and two single-channel leads are attached to them ($M=1$). Here $\langle S \rangle$ vs.~$\varepsilon$ follows the same behavior as for RGGs. That is, $\langle S \rangle\approx 1$ for $\varepsilon=0$, then $\langle S \rangle$ decreases for increasing $\varepsilon$ until it approximately vanishes at $\varepsilon = \varepsilon_{0}$, which marks the perfect coupling condition between the graphs and the leads. For further increasing $\varepsilon$, $\langle S \rangle$ grows steadily. Again, this behavior is well reproduced by Eq.~(\ref{eq:FittingS}), as shown in dashed lines in Fig.~(\ref{fig:BRGGsAvgS}). Moreover, as observed in Fig.~\ref{fig:BRGGsAvgS}, the perfect coupling condition depends on both the average degree $\langle k \rangle$ and the size of the sets composing the bipartite graph. Therefore, a detailed numerical analysis (not shown here) leads the coupling strength
\begin{equation}
\varepsilon_{0} \approx \left\{
\begin{array}{rl}
0.33 \langle k \rangle^{0.35} + 0.52 & \text{for } \mathrm{BRGG}(N/2), \\
0.56 \langle k \rangle^{0.57} + 0.42 & \text{for } \mathrm{BRGG}(N/5), \\
0.76 \langle k \rangle^{0.02} + 0.06 & \text{for } \mathrm{BRGG}(4N/5),
\end{array} \right.
\label{eq:BGepsilon0}
\end{equation}
such that $\langle S \rangle \approx 0$ is fulfilled. These expressions are used to fix the coupling strength used in the following calculations.


\subsection{Scattering matrix elements of BRGGs}

%
\begin{figure}
\centering
\includegraphics[width=0.99\columnwidth]{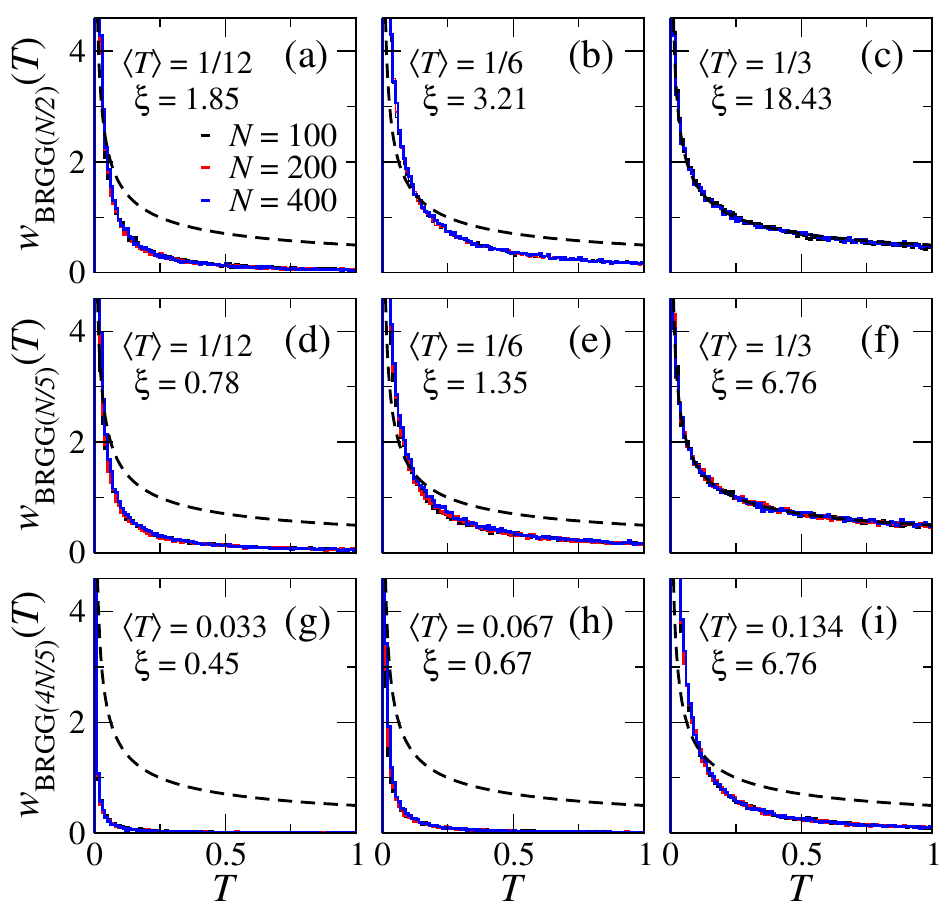}
\caption{(Color online) Transmission distribution $w(T)$ for tight-binding BRGGs in the two open-channels setup ($M=1$). The numerical results correspond to graph sizes of $N = 100$, 200, and 400, depicted in black, red, and blue histograms for (upper panels) BRGG($N/2$), (middle panels) BRGG($N/5$), and (lower panels) BRGG($4N/5$). Several fixed values of the average transmission are considered. 
The black dashed lines correspond to the RMT prediction~(\ref{eq:wTN1}).}
\label{fig:BRGGswTM1}
\end{figure}

The average of the squared modulus of the scattering matrix elements $\langle |S_{11}|^{2} \rangle$ and $\langle |S_{12}|^{2} \rangle$ is shown in Fig.~\ref{fig:BRGGsAvgSmnM1} for BRGG($N/2$), BRGG($N/5$), and BRGG($4N/5$) in left, middle, and right panels, respectively, as a function of (upper panels) $r$ and (lower panels) $\xi$. The scattering setup where two single-channel leads ($M = 1$) are attached to the BRGGs is considered with BRGGs of total sizes of $N = 100$, 200, and 400. As a function of the connection radius $r$, the $S$-matrix elements are size dependent. As in the RGGs case, a scaling parameter for BRGGs also takes the form given by expression~(\ref{eq:xi}) with $\alpha = 0.2166 \pm 0.0037$, $0.3375 \pm 0.0024$, and $0.3429 \pm 0.0371$ for BRGG($N/2$), BRGG($N/5$), and BRGG($4N/5$) respectively. Then, as observed in the lower panels of Fig.~\ref{fig:BRGGsAvgSmnM1}, both $\langle |S_{11}|^{2} \rangle$ and $\langle |S_{12}|^{2} \rangle$ are scaled by the parameter $\xi$; that is, the curves $\langle |S_{11}|^{2} \rangle$ and $\langle |S_{12}|^{2} \rangle$ versus $\xi$ fall on top of a universal curve (red continuous lines) given by~(\ref{eq:relSmnM}) with $\delta = 0.310$, 0.737, and 0.741 for BRGG($N/2$), BRGG($N/5$), and BRGG($4N/5$), respectively. 

Finally, in Fig.~\ref{fig:BRGGsAvgSmnM1} the black dashed lines correspond to the RMT prediction~(\ref{eq:Smn}) for the respective $S$-matrix elements. These RMT limit values are reached for mostly connected BRGG($N/2$) and BRGG($N/5$), meaning that for these scattering setups we observe the full insulating to metallic crossover by increasing the connectivity of the BRGGs. 
However, even complete BRGG($4N/5$) does not allow to reach the RMT predictions because here the leads are attached to vertices in the largest set. That is, in this scattering setup, the incoming waves are apparently not able to explore homogeneously the full graph and, as a consequence, the metallic regime is not reached.

%
\begin{figure}
\centering
\includegraphics[width=0.99\columnwidth]{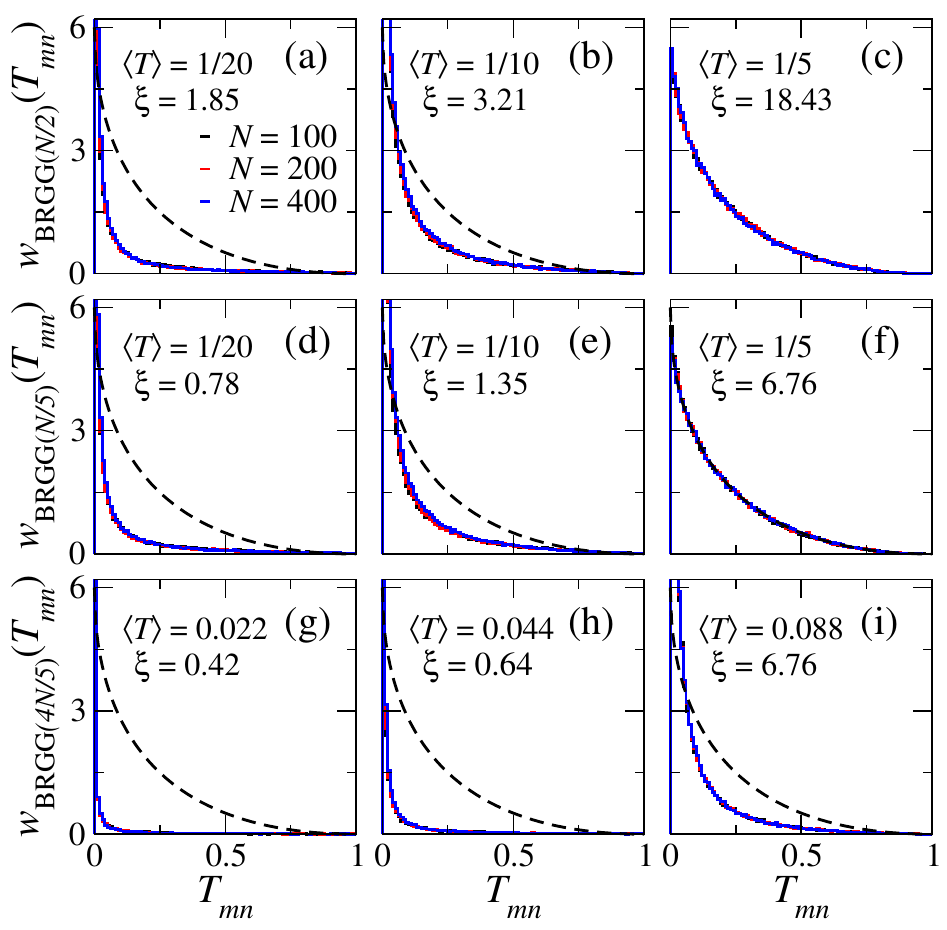}
\caption{(Color online) Channel-to-channel or mode-to-mode transmission distributions for tight-binding BRGGs with $M=2$ (four open-channels). The numerical results correspond to graph sizes of $N = 100$, 200, and 400, depicted in black, red, and blue histograms for (upper panels) BRGG($N/2$), (middle panels) BRGG($N/5$), and (lower panels) BRGG($4N/5$).
In each case, the average transmission is fixed as indicated in every panel. The black dashed lines correspond to the RMT prediction~(\ref{eq:wTnm}). The RMT limit is recovered for mostly connected BRGGs graphs, see panels (c) and (f), except for the case in panel (i).}
\label{fig:BRGGswTmnM2}
\end{figure}

Furthermore, in Fig.~\ref{fig:BRGGsAvgSmnM2M3} the average $S$-matrix elements $\langle |S_{mm}|^{2} \rangle$ (with $m = 1$, 2, 3, and 4) and $\langle |S_{mn}|^{2} \rangle$ (with $mn = 12$, 23, 34, and 41) as a function of $\xi$ are presented for BRGG($N/2$), BRGG($N/5$), and BRGG($4N/5$) in left, middle, and right panels, respectively. 
The total size of the BRGGs is $N = 200$. Figures~\ref{fig:BRGGsAvgSmnM2M3}(a-c) show the results when the BRGGs support four single-channel attached leads, $M = 2$, meanwhile Figs.~\ref{fig:BRGGsAvgSmnM2M3}(d-f) report the $M = 3$ case. In all panels, black dashed lines correspond to the respective RMT predictions given by Eq.~(\ref{eq:Smn}). A good agreement between the numerical results and the RMT prediction~(\ref{eq:Smn}) is obtained for mostly connected BRGGs for the cases BRGG($N/2$) and BRGG($N/5$). As in the $M = 1$ case (see Fig.~\ref{fig:BRGGsAvgSmnM1}), not even fully connected BRGG($4N/5$) reach the RMT prediction. The red continuous lines are fittings of expression~(\ref{eq:relSmnM}) to the numerical data where the fitting constants for the $M = 2$ (3) case are $\delta = 0.344$ (0.348), 0.818 (0.838), and 0.828 (0.880) for BRGG($N/2$), BRGG($N/5$), and BRGG($4N/5$), respectively. We stress that expression~(\ref{eq:relSmnM}) describes well all average $S$-matrix elements even in those cases where they do not approach the RMT predictions, as for BRGG($4N/5$).


\subsection{Transmission, shot noise power, and elastic enhancement factor for BRGGs}

%
\begin{figure}
\centering
\includegraphics[width=1.0\columnwidth]{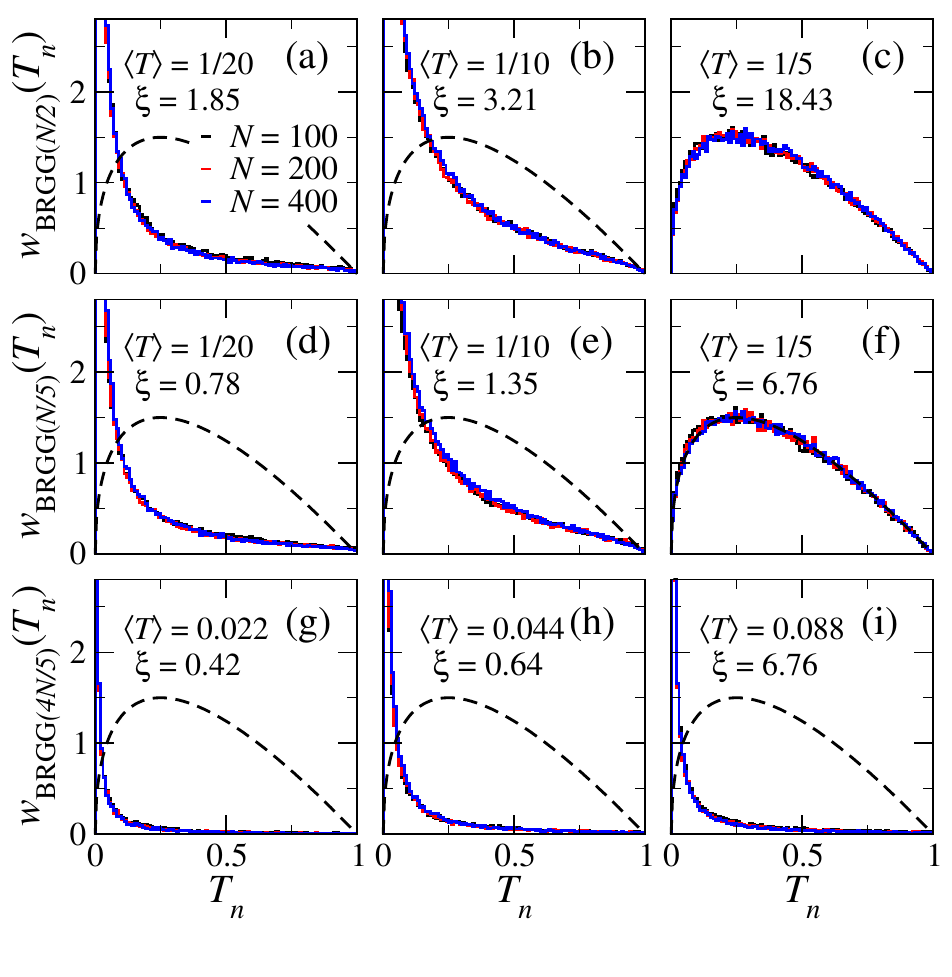}
\caption{Same graph parameters as in Fig.~\ref{fig:BRGGswTmnM2} but for the transmission distribution $w(T_{n})$. The black dashed lines correspond to the RMT prediction of Eq.~(\ref{eq:wTn}).}
\label{fig:BRGGswTnM2}
\end{figure}

Transmission distributions for BRGGs with two single-channel leads attached to them ($M = 1$) are depicted in Fig.~\ref{fig:BRGGswTM1}. The upper, middle, and lower panels show the results for BRGG($N/2$), BRGG($N/5$), and BRGG($4N/5$), respectively. For each case, BRGGs with different fixed average transmissions, or fixed $\xi$, and three graph sizes ($N = 100$, 200, and 400 in black, red, and blue histograms respectively) are considered. It is noticeable that the parameter $\xi$, see Eq.~(\ref{eq:xi}), properly scales the transmission distribution. Also, as the parameter $\xi$ increases, a smooth transition from insulating to metallic behavior is observed. For mostly connected graphs, a good agreement between the numerical results and the RMT prediction~(\ref{eq:wTN1}) (black dashed lines) is obtained, except for BRGG($4N/5$) as it happens for the $S$-matrix elements, see Figs.~\ref{fig:BRGGsAvgSmnM1}(c,f) and \ref{fig:BRGGsAvgSmnM2M3}(d,f).

\begin{figure}
\centering
\includegraphics[width=1.0\columnwidth]{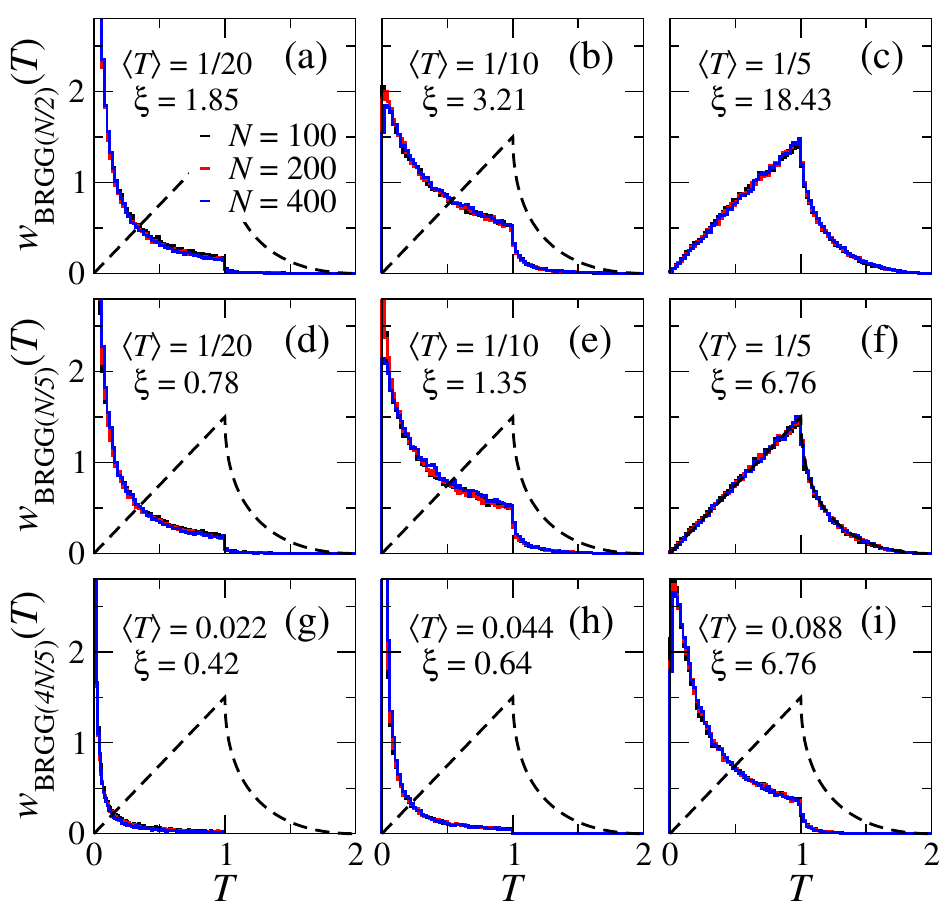}
\caption{Same graph parameters as in Figs.~\ref{fig:BRGGswTmnM2} and \ref{fig:BRGGswTnM2} but for the total transmission distribution $w(T)$. The black dashed lines correspond to the RMT prediction of Eq.~(\ref{eq:wTN2}).}
\label{fig:BRGGswTM2}
\end{figure}

In Figs.~\ref{fig:BRGGswTmnM2}, \ref{fig:BRGGswTnM2}, and \ref{fig:BRGGswTM2} the mode-to-mode transmission distributions $w(T_{mn})$, $w(T_{n})$, and $w(T)$, respectively, are depicted for BRGGs with four single-channel leads attached to them ($M = 2$). Upper, middle, and lower panels show the results for BRGG($N/2$), BRGG($N/5$), and BRGG($4N/5$), respectively. For each case, bipartite graphs with fixed average transmission, or equivalently fixed $\xi$ (as indicated in every panel) and three graph sizes $N$ (see panel (a) of each figure) are considered. The corresponding RMT predictions given by Eqs.~(\ref{eq:wTnm}-\ref{eq:wTN2}) are shown in black dashed lines. As observed, once $\xi$ is fixed, the distributions do not depend on the graph size, validating once again $\xi$ as scaling parameter. Also, a smooth transition from insulating to metallic behavior takes place as $\xi$ increases where a good agreement with the RMT predictions are obtained for mostly connected graphs, except as expected, for the case of BRGG($4N/5$).

%

\begin{figure}
\centering
\includegraphics[width=1.0\columnwidth]{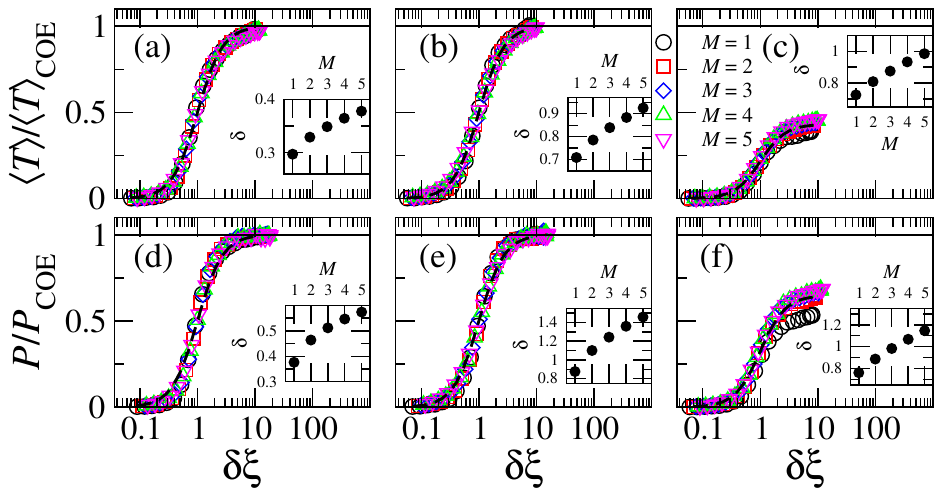}
\caption{(Color online) (Upper panels) Average transmission $\langle T \rangle$ and (lower panels) shot noise power $P$ (normalized to their respective RMT limit values given by Eqs.~(\ref{eq:AvgT}) and (\ref{eq:P}), respectively) as a function of $\delta \xi$ for tight-binding BRGGs with $N = 200$ and (left panels) BRGG($N/2$), (middle panels) BRGG($N/5$), and (right panels) BRGG($4N/5$). $2M$ single-channel leads are attached to the graphs with $M = 1, \ldots, 5$. The black dashed lines correspond to the fitting of Eq.~(\ref{eq:Xxi}) to the numerical data. The value of the fitting parameter $\delta$ for each case is shown in the corresponding insets. The black full lines are shown to guide the eye.}
\label{fig:BRGGsTP}
\end{figure}

In Fig.~\ref{fig:BRGGsTP} the average transmission and the shot noise power (normalized by their corresponding RMT limit values given in Eqs.~(\ref{eq:AvgT}) and (\ref{eq:P}), respectively) are plotted as a function of $\delta \xi$ for tight-binding BRGGs of size $N = 200$. Here, the BRGGs support $2M = 2, 4, \ldots, 10$ open channels. Upper panels show the results for the average transmission of BRGG($N/2$), BRGG($N/5$), and BRGG($4N/5$), in panels (a), (b), and (c) respectively. Lower panels show the respective results for the shot noise power. The symbols are obtained from numerical simulations while the black continuous lines at 1 are shown to guide the eye. The black dashed lines correspond to fittings of expression~(\ref{eq:Xxi}) to the numerical data. The respective fitting parameters $\delta$ are shown in the corresponding insets. As observed in Fig.~\ref{fig:BRGGsTP}, expression~(\ref{eq:Xxi}) correctly describes the numerical data for both the average transmission as well as the shot noise power, even for the cases of BRGG($4N/5$) which do not reach the RMT limits. Error bars are not shown since they are smaller than the symbols.

\begin{figure}
\centering
\includegraphics[width=1.0\columnwidth]{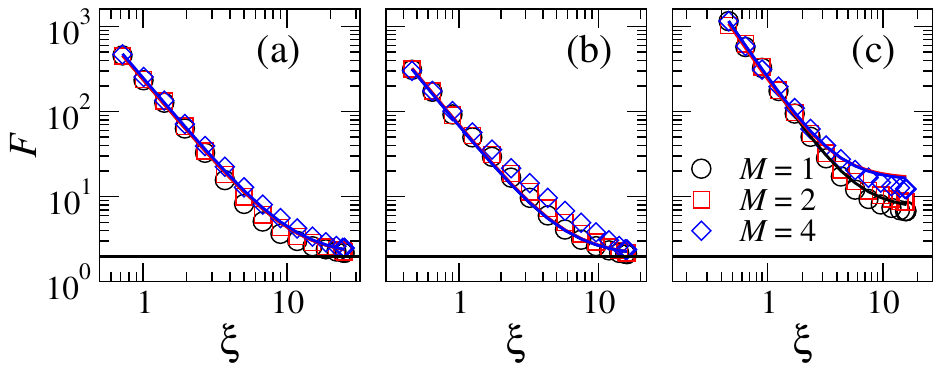}
\caption{(Color online) Elastic enhancement factor $F$ as a function of $\xi$ for tight-binding BRGGs with $N = 200$ and (a) BRGG($N/2$), (b) BRGG($N/5$), and (c) BRGG($4N/5$). $2M$ single-channel leads are attached to the graphs with $M = 1$ (black circles), 2 (red squares), and 4 (blue diamonds) open channels. The black (red, blue) continuous line corresponds to fitting of Eq.~(\ref{eq:FitF}) to the numerical data for the $M = 1$ (2, 4) case with fitting constant a) $c = 236.28\, (235.14, 246.96)$, b) $c = 66.97\, (68.40, 66.40)$, and c) $c = 243.78\, (225.88, 242.32)$, respectively. The black horizontal line in each panel corresponds to the RMT limit value of $F = 2$. This RMT limit is only reached for mostly connected BRGG($N/2$) and BRGG($N/5$).}
\label{fig:BRGGsF}
\end{figure}

To finalize with the statistical analysis of the scattering and transport properties of tight-binding BRGGs, in Fig.~\ref{fig:BRGGsF} the elastic enhancement factor $F$ as a function of $\xi$ is shown for BRGG($N/2$), BRGG($N/5$), and BRGG($4N/5$) having $N = 200$ vertices. 
From this figure, it is observed that for any number of open channels $2M$, $F$ decreases as a function of $\xi$ and approaches smoothly, for large $\xi$, the RMT limit value of 2 for the BRGG($N/2$) and BRGG($N/5$) cases. As occurs for the other scattering and transport properties analyzed in this work, the RMT limit of $F=2$ is not reached for BRGG($4N/5$). Nevertheless, the behavior of $F$ in all cases is well captured by the expression~(\ref{eq:FitF}).

%
%


\section{Conclusions}
\label{sec:Conclusions}

In this paper, an extensive numerical analysis of the scattering and transport properties of tight-binding random geometric graphs (RGGs) and bipartite random geometric graphs (BRGGs) has been presented. These properties correspond to the scattering matrix elements, the transmission, the mode-to-mode transmission distributions, the shot noise power, and the elastic enhancement factor. The RGGs are characterized by the total number of vertices $N$ and the connection radius $r$. The BRGGs, on the other hand, are characterized by $N$ and $r$, but also by the number of vertices ($s$ and $N-s$) of the two disjoint sets (A and B) forming the bipartition. Specifically, in the case of BRGGs, we chose a scattering setup where leads are attached to vertices in set A, which can be the smallest or the largest set.

The scattering and transport properties of both RGGs and BRGGs show a universal behavior once those properties are properly scaled. The scaling parameter $\xi$, see Eq.~(\ref{eq:xi}), is proportional to the average degree but shows a further dependence on the graph size as a power law, $\xi\sim N^{-\alpha}$. Here we found that $\alpha=0.26\pm 0.018$ for RGGs, irrespective of the number of attached leads. While for BRGGs, $\alpha$ could be even larger depending on the size ratio between the sets composing the bipartition. 

At this point it becomes relevant to recall that in Ref.~\cite{Martinez2013} it was shown that the average degree serves as the scaling parameter of scattering and transport properties of Erd\"os-Renyi graphs (ERGs). This, in our terms, means that as well as for RGGs and BRGGs the scattering and transport properties of ERGs accept a scaling parameter of the form given by Eq.~(\ref{eq:xi}) but with $\alpha= 0$. However, a detailed analysis (not shown here) reveals that $\alpha\ne 0$. In fact we have found that $\alpha = 0.075 \pm 0.0029$ for ERGs; indeed a small value of $\alpha$ but clearly measurable.

%

Furthermore, by leveraging the equivalence that exists between the adjacency matrix describing both RGGs and BRGGs and the tight-binding Hamiltonian describing the corresponding electronic structures, the numerical results have been contrasted with random-matrix theory (RMT) predictions for electronic transport under the perfect coupling regime between the graphs and the leads. Also, the scattering and transport properties studied here show a smooth transition from an insulating to a metallic regime (described by the RMT) for increasing graph connectivity. Indeed, a good correspondence between the numerical results and the RMT predictions has been observed when the graphs are mostly connected, except for BRGGs in a specific scattering setup. Even though here we have reported only one case of BRGGs where the metallic regime is not reached, we have verified that this always happens in setups with leads attached to vertices belonging to the largest set. Moreover, the larger the set is, the larger the difference between the scattering and transport properties and the corresponding RMT predictions is.

Additionally, scaled curves like those shown in Figs.~2(c),~3,~9, and~10 allow to properly define the transport regimes:
for both RGGs and BRGGs the insulating (metallic) regime is observed when $\xi<1$ ($\xi>10$), while the insulating-to-metallic transition takes place in the interval $1<\xi<10$. Of course, BRGGs that do not reach the metallic regime should be excluded from this statement. 

Finally, it is worth mentioning that the aforementioned equivalence that exists between the adjacency matrix describing random graphs and the tight-binding Hamiltonian describing electronic systems may assist to deepen in the understanding of random graphs by leveraging well known results from RMT.


\acknowledgments

A.M.M-A.\ acknowledges financial support from CONAHCyT under the program ``Estancias Posdoctorales por M\'exico 2022".
J.A.M.-B. thanks support from CONACyT (Grant No. 286633), CONACyT-Fronteras (Grant No. 425854), 
and VIEP-BUAP (Grant No. 100405811-VIEP2023), Mexico.







\begin{thebibliography}{50}

\bibitem{Strogatz2001}
S. H. Strogatz,
Nature \textbf{410}, 268 (2001)

\bibitem{Dorogovtsev2001}
S. N. Dorogovtsev and J. F. F. Mendes,
Adv. Phys. \textbf{51}, 1079 (2001)

\bibitem{Albert2002}
R. Albert and A. L. Barab\'asi,
Rev. Mod. Phys. \textbf{74}, 47 (2002)

\bibitem{Newman2003}
M. E. J. Newman,
SIAM Rev. \textbf{45}, 167 (2003)

\bibitem{Boccalettia2006}
S. Boccalettia, V. Latora, Y. Moreno, M. Chavez, D.-U. Hwang,
Phys. Rep. \textbf{424}, 175 (2006)

\bibitem{DurretBook}
R. Durret,
\emph{Random Graph Dynamics} (Cambridge University Press, UK, 2007)

\bibitem{LatoraBook}
V. Latora, V. Nicosia, and G. Russo, 
\emph{Complex Networks Principles, Methods and Applications} (Cambridge University Press, UK, 2017)

\bibitem{EstradaBook}
E. Estrada, \emph{The Structure of Complex Networks: Theory and Applications} (Oxford University Press, New York, 2011).

\bibitem{PenroseBook}
M. Penrose, \emph{Random Geometric Graphs} (Oxford University Press, New York, 2003)

\bibitem{Erdos1959}
P. Erd\"os and A. R\'enyi,
Publicationes Mathematicae \textbf{6}, 290 (1959)

\bibitem{Chung1997}
F. R. K. Chung,
\emph{Spectral Graph Theory} (American Mathematical Society, Providence, R.I. 1997)

\bibitem{Bianconi2001}
G. Bianconi and A.-L. Barabsi.
Phys. Rev. Lett. \textbf{86}, 5632 (2001)

\bibitem{Dorogovtsev2008}
S. N. Dorogovtsev and A. V. Goltsev,
Rev. Mod. Phys. \textbf{80}, 1275 (2008)

\bibitem{Estrada2012}
E. Estrada, N. Hatano, and M. Benzi,
Phys. Rep. \textbf{514}, 89 (2012)

\bibitem{Holovatch2017}
Y. Holovatch, R. Kenna, and S. Thurner,
Eur. J. Phys. \textbf{38}, 023002 (2017)

\bibitem{Biamonte2019}
J. Biamonte, M. Faccin, and M. De Domenico,
Comms. Phys. \textbf{2}, 53 (2019)

\bibitem{Cimini2019}
G. Cimini, T. Squartini, F. Saracco, D. Garlaschelli, A. Gabrielli, and G. Caldarelli,
Nat. Rev. Phys. \textbf{1}, 58 (2019)

\bibitem{Jusup2022}
M. Jusup, et. al.
Phys. Rep. \textbf{948}, 1 (2022)

\bibitem{Lehn2013}
J.-M. Lehn,
Angew. Chem. Int. Ed. \textbf{52}, 2 (2013)

\bibitem{Rappoport2014}
D. Rappoport, C. J. Galvin, D. Y. Zubarev, and A. Aspuru-Guzik,
J. Chem. Theory Comput. \textbf{10}, 897 (2014)

\bibitem{Ozkanlar2014}
A. Ozkanlar and A. E. Clark,
J. Compt. Chem. \textbf{35}, 495 (2014)

\bibitem{Kohn1999}
K. W. Kohn,
Mol. Biol. Cell \textbf{10}, 2703 (1999)

\bibitem{Hartwell1999}
L. H. Hartwell, J. J. Hopfield, S. Leibler, and A. W. Murray,
Nature \textbf{402}, C47 (1999)

\bibitem{Bhalla1999}
U. S. Bhalla and R. Iyengar,
Science \textbf{283}, 381 (1999)

\bibitem{Wong2006}
L. H. Wong, P. Pattison, and G. Robins,
Physica A \textbf{360}, 99 (2006)

\bibitem{Barabasi1999}
A. L. Barab\'asi and R. Albert,
Science \textbf{286}, 509 (1999)

\bibitem{Williams2000}
R. J. Williams and N. D. Mart\'inez, 
Nature \textbf{404}, 180 (2000)

\bibitem{Abello1998}
J. Abello, A. Buchsbaum, and J. A. Westbrook,
Lect. Notes Comput. Sci. \textbf{1461}, 332 (1998).

\bibitem{Broder2000}
A. Broder, et al.
Comput. Netw. \textbf{33}, 309 (2000)

\bibitem{Urban2001}
D. Urban andT. Keitt,
Ecology \textbf{82}, 1205 (2001)

\bibitem{Lopez2005}
E. L\'opez, S. V. Buldyrev, S. Havlin, and H. E. Stanley,
Phys. Rev. Lett. \textbf{94}, 248701 (2005)

\bibitem{Lopez2006}
E. L\'opez, S. Carmib, S. Havlin, S. V. Buldyrev, H. E. Stanley,
Physica D \textbf{224}, 69 (2006)

\bibitem{Wu2006}
Z. Wu, L. A. Braunstein, S. Havlin, and H. E. Stanley,
Phys. Rev. Lett. \textbf{96}, 148702 (2006)

\bibitem{Gallos2007}
L. K. Gallos, C. Song, S. Havlin, and H. A. Makse,
PNAS \textbf{104}, 7746 (2007)

\bibitem{Candia2007}
J. Candia, P. E. Parris, V. M. Kenkre,
J. Stat. Phys. \textbf{129}, 323 (2007)

\bibitem{Ramasco2007}
J. J. Ramasco and G. Gon\c{c}alves,
Phys. Rev. E \textbf{76}, 066106 (2007) 

\bibitem{Li2007}
G. Li, L. A. Braunstein, S. V. Buldyrev, S. Havlin, and H. E. Stanley,
Phys. Rev. E \textbf{75}, 045103(R) (2007)

\bibitem{Nicolaides2010}
C. Nicolaides, L. Cueto-Felgueroso, and R. Juanes
Phys. Rev. E textbf{82}, 055101(R) (2010)

\bibitem{Mulken2011}
O. M\"ulken and A. Blumen,
Phys. Rep. \textbf{502}, 37 (2011)

\bibitem{Xiong2018}
K. Xiong, C. Zeng, and Z. Liu,
Nonlinear Dyn. \textbf{94}, 3067 (2018)

\bibitem{Zhu2000}
C.-P. Zhu and S.-J. Xiong,
Phys. Rev. B \textbf{62}, 14780 (2000)

\bibitem{Giraud2005}
O. Giraud, B. Georgeot, and D. L. Shepelyansky,
Phys. Rev. E \textbf{72}, 036203 (2005)

\bibitem{Sade2005}
M. Sade, T. Kalisky, S. Havlin, and R. Berkovits,
Phys. Rev. E \textbf{72}, 066123 (2005)

\bibitem{Slanina2012}
F. Slanina,
Eur. Phys. J. B \textbf{85}, 361 (2012)

\bibitem{Bandyopadhyay2007}
J. N. Bandyopadhyay and S. Jalan,
Phys. Rev. E \textbf{76}, 026109 (2007)

\bibitem{Jalan2007}
S. Jalan and J. N. Bandyopadhyay,
Phys. Rev. E \textbf{76}, 046107 (2007)

\bibitem{Jalan2009}
S. Jalan,
Phys. Rev. E \textbf{80}, 046101 (2009)

\bibitem{deCarvalho2009}
J. X. deCarvalho, S. Jalan, and M. S. Hussein,
Phys. Rev. E \textbf{79}, 056222 (2009)

\bibitem{Jalan2010}
S. Jalan, N. Solymosi, G. Vattay, and B. Li,
Phys. Rev. E \textbf{81}, 046118 (2010)

\bibitem{Zhu2008}
G. Zhu, H. Yang, C. Yin, and B. Li,
Phys. Rev. E \textbf{77}, 066113 (2008)

\bibitem{Jahnke2008}
L. Jahnke, J.W. Kantelhardt, R. Berkovits, and S. Havlin, 
Phys. Rev. Lett. \textbf{101}, 175702 (2008)

\bibitem{Goh2001}
K.-I. Goh, B. Kahng, and D. Kim,
Phys. Rev. E \textbf{64}, 051903 (2001)

\bibitem{Rodgers2005}
G. J. Rodgers, K. Austin, B. Kahng, and D. Kim,
J. Phys. A: Math. Gen. \textbf{38}, 9431 (2005)

\bibitem{Nagao2008}
T. Nagao and G. J. Rodgers,
J. Phys. A: Math. Theor. \textbf{41}, 265002 (2008)

\bibitem{Georgeot2010}
B. Georgeot, O. Giraud, and D. L. Shepelyansky, 
Phys. Rev. E \textbf{81}, 056109 (2010)

\bibitem{Jalan2011}
S. Jalan, G. Zhu, and B. Li,
Phys. Rev. E \textbf{84}, 046107 (2011)

\bibitem{Gong2006}
L. Gong and P. Tong,
Phys. Rev. E \textbf{74}, 056103 (2006)

\bibitem{Farkas2002}
I. Farkas, I. Der\'enyi, H. Jeong, Z. N\'eda, Z. N. Oltvai, E. Ravasz, A. Schubert, A. L. Barab\'asi, and T. Vicsek,
Physica A \textbf{314}, 25 (2002)

\bibitem{Dorogovtsev2003}
S. N. Dorogovtsev, A. V. Goltsev, J. F. F. Mendes, and A. N. Samukhin,
Phys. Rev. E \textbf{68}, 046109 (2003)

\bibitem{Martinez2019}
C. T. Mart\'inez-Mart\'inez, J. A. M\'endez-Berm\'udez, Y. Moreno, J. J. Pineda-Pineda, J. M. Sigarreta,
Chaos, Solitons, and Fractals X \textbf{3}, 100021  (2019)

\bibitem{Martinez2022}
C. T. Mart\'inez-Mart\'inez, J. A. M\'endez-Berm\'udez, F. A. Rodrigues, and E. Estrada,
Phys. Rev. E \textbf{105}, 034304 (2022)

\bibitem{Rodgers1988}
G. J. Rodgers and A. J. Bray,
Phys. Rev. B \textbf{37}, 3557 (1988).

\bibitem{Fyodorov1991a}
Y. V. Fyodorov and A. D. Mirlin,
J. Phys. A: Math. Gen. \textbf{24}, 2219 (1991)

\bibitem{Fyodorov1991b}
Y. V. Fyodorov and A. D. Mirlin,
Phys. Rev. Lett. \textbf{67}, 2049 (1991)

\bibitem{Mirlin1991}
A. D. Mirlin and Y. V. Fyodorov,
J. Phys. A: Math. Gen. \textbf{24}, 2273 (1991)

\bibitem{Evangelou1992a}
S. N. Evangelou and E. N. Economou, 
Phys. Rev. Lett. \textbf{68}, 361 (1992)

\bibitem{Evangelou1992b}
S. N. Evangelou,
J. Stat. Phys. \textbf{69}, 361 (1992)

\bibitem{Jackson2001}
A. D. Jackson, C. Mejia-Monasterio, T. Rupp, M. Saltzer, and T. Wilke,
Nucl. Phys. A \textbf{687}, 405 (2001)

\bibitem{Barthelemy2011}
M. Barth\'elemy,
Phys. Rep. \textbf{499}, 1 (2011)

\bibitem{Gilbert1959}
E. N. Gilbert, 
Ann. Math. Stat. \textbf{30}, 1141 (1959)

\bibitem{Dall2002}
J. Dall and M. Christensen,
Phys. Rev. E \textbf{66}, 016121 (2002).

\bibitem{Stegehuis2022}
C. Stegehuis, L. Weedage,
Physica A \textbf{586}, 126460 (2022)

\bibitem{Allen2018}
A. Allen-Perkins,
Phys. Rev. E \textbf{98}, 032310 (2018).

\bibitem{Estrada2015}
E. Estrada and M. Sheerin,
Phys. Rev. E \textbf{91}, 042805 (2015).

\bibitem{Pratt2018}
P. Pratt, C. P. Dettmann, and W. H. Chin,
Phys. Rev. E \textbf{98}, 052310 (2018).

\bibitem{Estrada2017}
E. Estrada,
Phys. Rev. E \textbf{96}, 022314 (2017).

\bibitem{Waxman1988}
B. M. Waxman,
IEEE J. Sel. Areas Commun. \textbf{6}, 1617 (1988)

\bibitem{Alonso2018}
L. Alonso, J. A. M\'endez-Berm\'udez, A. Gonz\'alez-Mel\'endrez, and Y. Moreno,
J. Complex Netw. \textbf{6}, 753 (2018)

\bibitem{Aguilar2020}
R. Aguilar-S\'anchez, J. A. M\'endez-Berm\'udez, F. A. Rodrigues, and J. M. Sigarreta,
Phys. Rev. E \textbf{102}, 042306 (2020)

\bibitem{PM23}
K. Peralta-Mart\'inez and J. A. M\'endez-Berm\'udez,
J. Phys. Complex. \textbf{4}, 015002 (2023)

\bibitem{Beenakker1997}
C. W. J. Beenakker, 
Rev. Mod. Phys. \textbf{69}, 731 (1997)

\bibitem{MelloBook}
P. A. Mello and N. Kumar, \emph{Quantum Transport in Mesoscopic Systems: Complexity and Statistical Fluctuations} (Oxford Universtity Press, New York, 2005).

\bibitem{Martinez2013}
A. J. Mart\'inez-Mendoza, A. Alcazar-L\'opez, and J. A. M\'endez-Berm\'udez,
Phys. Rev. E \textbf{88}, 122126 (2013)

\bibitem{Verbaarschot1985}
J. J. M. Verbaarschot, H. A. Weidenm\"uller, and M. R. Zirnbauer,
Phys. Rep. \textbf{129}, 367 (1985)

\bibitem{vanLangen1996}
S. A. van Langen, P. W. Brouwer, and C. W. J. Beenakker
Phys. Rev. E \textbf{53}, R1344(R) (1996)

\bibitem{Pereyra1983}
P. Pereyra and P A Mello,
J. Phys. A: Math. Gen. \textbf{16}, 237 (1983)

\bibitem{Savin2006}
D. V. Savin and H.-J. Sommers, 
Phys. Rev. B \textbf{73}, 081307(R) (2006)

\bibitem{Harney1986}
H. L. Harney, A. Richter, and H. A. Weidenm\"uller,
Rev. Mod. Phys. \textbf{58}, 607 (1986)

\bibitem{Mendez2010}
J. A. M\'endez-Berm\'udez, V. A. Gopar, and I. Varga,
Phys. Rev. B \textbf{82}, 125106 (2010)

\bibitem{AMMA2023}
A. M. Mart\'inez-Arg\"uello, M. Carrera-N\'u\~nez, and J. A. M\'endez-Berm\'udez,
Phys. Rev. E \textbf{107}, 024139 (2023)

\end{thebibliography}


\end{document}